\documentclass[aps,prd,twocolumn,showpacs,superscriptaddress,nofootinbib]{revtex4-2}

\usepackage{amsmath,amssymb,amsfonts}
\usepackage{bm}
\usepackage{graphicx}
\usepackage{tikz}
\usepackage{multirow}
\usepackage{hyperref}
\usepackage{slashed}
\usepackage{color}
\usepackage{comment}
\usepackage{mathtools}
\usepackage{physics}
\usepackage{tensor}
\usepackage{float}
\usepackage{tikz}
\usepackage{pgfplots}
\pgfplotsset{compat=1.17}
\usepackage{graphicx}
\usepackage{dblfloatfix}
\usepackage[utf8]{inputenc}
\usepackage[T1]{fontenc}
\usepackage{amsmath,amssymb}
\usepackage{physics}
\usepackage{slashed}

\begin{document}

\title{
Hybrid Type-I and Type-II Leptogenesis in Minimal Renormalizable
$\mathrm{SO}(10)$
}

\author{Gayatri Ghosh}
\email{gayatrighsh@gmail.com}

\affiliation{
Department of Physics,
Cachar College, Assam University
}

\date{\today}


\begin{abstract}

We investigate flavoured hybrid leptogenesis in a minimal renormalizable
$\mathrm{SO}(10)$ framework in which both Type-I and Type-II seesaw
interactions contribute to the generation of the cosmological baryon
asymmetry. We consider a regime in which heavy Majorana neutrinos and
electroweak scalar triplets are independently quasi-degenerate, allowing
for resonant enhancement within each sector, while additional
CP-violating contributions arise from loop-induced interference between
fermionic and scalar interactions. These hybrid contributions vanish in
the absence of either sector and provide an additional source of
CP violation beyond the conventional single-sector mechanisms.

The evolution of the lepton asymmetry is analysed using a
flavour-covariant density-matrix formalism that consistently incorporates
flavour coherence, spectator effects, and washout processes. We show
that the observed baryon asymmetry,
\[
Y_B \simeq 8.7\times10^{-11},
\]
can be reproduced within representative regions of the parameter space
for heavy mass scales of
$10^{10}$--$10^{12}\,\mathrm{GeV}$ and perturbative Yukawa couplings.

To characterize the flavour structure of the hybrid interference terms,
we introduce the scalar combination
\[
J_{\rm hybrid}
=
{\rm Im}\!\left[
\mu\,{\rm Tr}
\!\left(
fY_\nu^\dagger Y_\nu
\right)
\right],
\]
which is employed throughout the analysis as a compact parametrization
of the flavour contractions entering the hybrid interference amplitudes
in the flavour basis adopted in this work. We discuss the role of this
combination in the hybrid amplitudes without assuming that it provides a
complete characterization of CP violation in the general framework.

The same interactions responsible for baryogenesis may also induce
charged-lepton flavour violation and electric dipole moments, providing
a phenomenological connection between neutrino mass generation, grand
unification, and future precision searches for CP violation.

\end{abstract}


\maketitle

\section{Introduction}

The origin of the baryon asymmetry of the Universe (BAU),
\[
Y_B=(8.70\pm0.06)\times10^{-11},
\]
remains one of the outstanding open problems in particle physics and
cosmology \cite{Planck2018,PDG2024}. Although the Standard Model (SM)
contains all three Sakharov conditions in principle, the amount of CP
violation and the nature of the electroweak phase transition are
insufficient to account for the observed asymmetry. This strongly
motivates the existence of new physics beyond the electroweak scale.

Among the proposed mechanisms, leptogenesis provides a particularly
attractive framework in which a primordial lepton asymmetry is generated
at high energies and subsequently converted into the observed baryon
asymmetry through electroweak sphaleron processes
\cite{FukugitaYanagida,DavidsonNardiNir,GiudiceEtAl}. An important
feature of this scenario is its intimate connection with neutrino mass
generation through the seesaw mechanism, thereby relating the origin of
the BAU to physics responsible for neutrino masses and flavour.

Canonical realizations of leptogenesis are generally classified according
to the source of lepton-number violation. In Type-I leptogenesis, the
asymmetry originates from the out-of-equilibrium decays of heavy
Majorana neutrinos
\cite{CoviRouletVissani,BuchmullerPlumacher,AbadaEtAlFlavour},
whereas in Type-II leptogenesis it is generated through the decays of
electroweak scalar triplets carrying lepton number
\cite{HambyeTypeII,AristizabalSierraHambye}. Both mechanisms can account
for the observed BAU over suitable regions of parameter space, although
successful leptogenesis typically requires favourable flavour
structures, sufficiently large CP asymmetries, or appropriate heavy-mass
spectra. Resonant enhancement of the CP asymmetry may occur when
particles with identical quantum numbers become quasi-degenerate,
leading to enhanced self-energy contributions within either the
heavy-neutrino or scalar-triplet sector separately
\cite{PilaftsisResonant,DevReviewResonant,BlanchetReview}.

Grand unified theories provide a natural framework in which multiple
sources of lepton-number violation coexist. In particular, minimal
renormalizable $\mathrm{SO}(10)$ models
\cite{GeorgiSO10,FritzschMinkowski,SenjanovicReview}
accommodate all fermions of a given generation within a single spinorial
representation and naturally realize both Type-I and Type-II seesaw
mechanisms. After symmetry breaking, the Higgs sector contains
right-handed neutrino mass terms together with electroweak scalar
triplets, allowing both seesaw contributions to arise from the same
underlying unified framework
\cite{LazaridesShafiWetterich,BajcSenjanovic,AltarelliFeruglioSO10}.
Depending on the structure of the Yukawa sector and scalar potential,
quasi-degenerate spectra may occur independently within the
heavy-neutrino and scalar-triplet sectors, providing phenomenologically
interesting regions for leptogenesis.

The interplay between Type-I and Type-II contributions, including
interference effects between the two sectors, has been investigated in
various contexts
\cite{BhupalDev:2022xyz}. Most previous studies, however, have been
formulated within generic seesaw frameworks rather than within a minimal
renormalizable grand unified theory. In the present work we investigate
hybrid leptogenesis in a renormalizable $\mathrm{SO}(10)$ framework,
where both heavy Majorana neutrinos and scalar triplets originate from
the same unified theory. We focus on parameter regions in which
quasi-degenerate spectra may arise independently within the
heavy-neutrino and scalar-triplet sectors. While resonant enhancement,
when present, is associated with quasi-degenerate states belonging to a
single sector, loop-induced interference between fermionic and scalar
interactions provides additional contributions to the CP asymmetry that
arise only when both sectors participate in the dynamics.

The flavour structure of CP violation in neutrino mass models has been
widely investigated using weak-basis invariants and related
rephasing-invariant combinations of couplings
\cite{Jarlskog1985,BrancoLavouraRebelo,BrancoBook,
DavidsonNardiNir,DevReviewResonant1}. Such quantities provide useful
tools for identifying physical CP-violating parameters independently of
particular field parametrizations. Rather than attempting a complete
classification of all possible invariant structures, the present work
focuses on the flavour structure relevant to the hybrid interference
contributions arising in the framework considered here.

To facilitate this discussion, we introduce the scalar combination
\[
J_{\rm hybrid}
=
{\rm Im}\!\left[
\mu\,{\rm Tr}
\left(
fY_\nu^\dagger Y_\nu
\right)
\right],
\]
which is employed throughout this work as a compact parametrization of
one class of flavour contractions appearing in the hybrid interference
amplitudes within the flavour basis adopted for the present analysis.
Its role in organizing the flavour structure of the hybrid interference
contributions is discussed in Sec.~III.

The generation of the lepton asymmetry is studied using
flavour-covariant kinetic equations within a density-matrix formalism,
which consistently incorporates flavour coherence, washout effects and
spectator processes. We explore representative regions of parameter
space characterized by heavy mass scales of
$10^{10}$--$10^{12}\,\mathrm{GeV}$ and perturbative Yukawa couplings,
showing that the observed baryon asymmetry can be reproduced while
remaining compatible with the phenomenological constraints considered
throughout this work. We also discuss possible implications for
charged-lepton flavour violation, including
$\mu\rightarrow e\gamma$, and electric dipole moments of the electron
and neutron
\cite{CiriglianoEDM1,ACME2018,ACME2023}.

The paper is organized as follows. In Sec.~II we introduce the
renormalizable $\mathrm{SO}(10)$ framework and the corresponding hybrid
seesaw structure. Section~III discusses the CP asymmetries and the
flavour structure of the hybrid interference contributions. The
numerical framework and benchmark scenarios are presented in Sec.~IV,
followed by the numerical results in Sec.~V. Our conclusions are given
in Sec.~VI.

\section{SO(10) Framework and Hybrid Seesaw Structure}

We outline the minimal renormalizable $\mathrm{SO}(10)$ framework that
forms the basis of the hybrid leptogenesis scenario investigated in this
work. Here ``minimal'' denotes the renormalizable model containing only
the Higgs representations $10_H$ and $\overline{126}_H$, without the
introduction of additional scalar multiplets or higher-dimensional
operators. Such constructions provide a predictive and ultraviolet-complete
framework for fermion mass generation, gauge unification, and neutrino
physics while maintaining a comparatively constrained parameter space at
high energies
\cite{AltarelliFeruglioSO10,FukuyamaReview2021,SenjanovicReview,DiLuzio2021,Babu2020SO10Fit}.

In this framework, each generation of Standard Model fermions is embedded
into a single spinorial representation $16_F$, which naturally contains
the right-handed neutrino required for the seesaw mechanism and ensures
anomaly cancellation. The renormalizable Yukawa sector is constructed
from three copies of $16_F$ coupled to the Higgs multiplets
\[
10_H \oplus \overline{126}_H,
\]
which constitute the minimal scalar content necessary to generate charged
fermion masses, Majorana neutrino masses, and the scalar-triplet
interactions relevant for the present analysis
\cite{BajcSenjanovic,GohMohapatraNg,GrimusKuhbock}. Owing to the
$\mathrm{SO}(10)$ contraction $16_F\,16_F\,H$, the Yukawa coupling
matrices are complex symmetric in flavour space.

After spontaneous symmetry breaking to the Standard Model gauge group,
the Higgs multiplets decompose into electroweak doublets, singlets, and
triplets. In particular, the representation $\overline{126}_H$
contains an $SU(2)_L$ scalar triplet,
\[
T \sim (1,3,1),
\]
which contributes to neutrino mass generation through the Type-II
seesaw mechanism. At the same time, the singlet components generate
Majorana masses for the right-handed neutrinos, giving rise to the
Type-I seesaw. Consequently, both seesaw mechanisms arise naturally
within the same renormalizable $\mathrm{SO}(10)$ framework and may
contribute simultaneously to the leptogenesis dynamics.
\begin{figure}[t]
\centering
\includegraphics[width=0.92\linewidth]{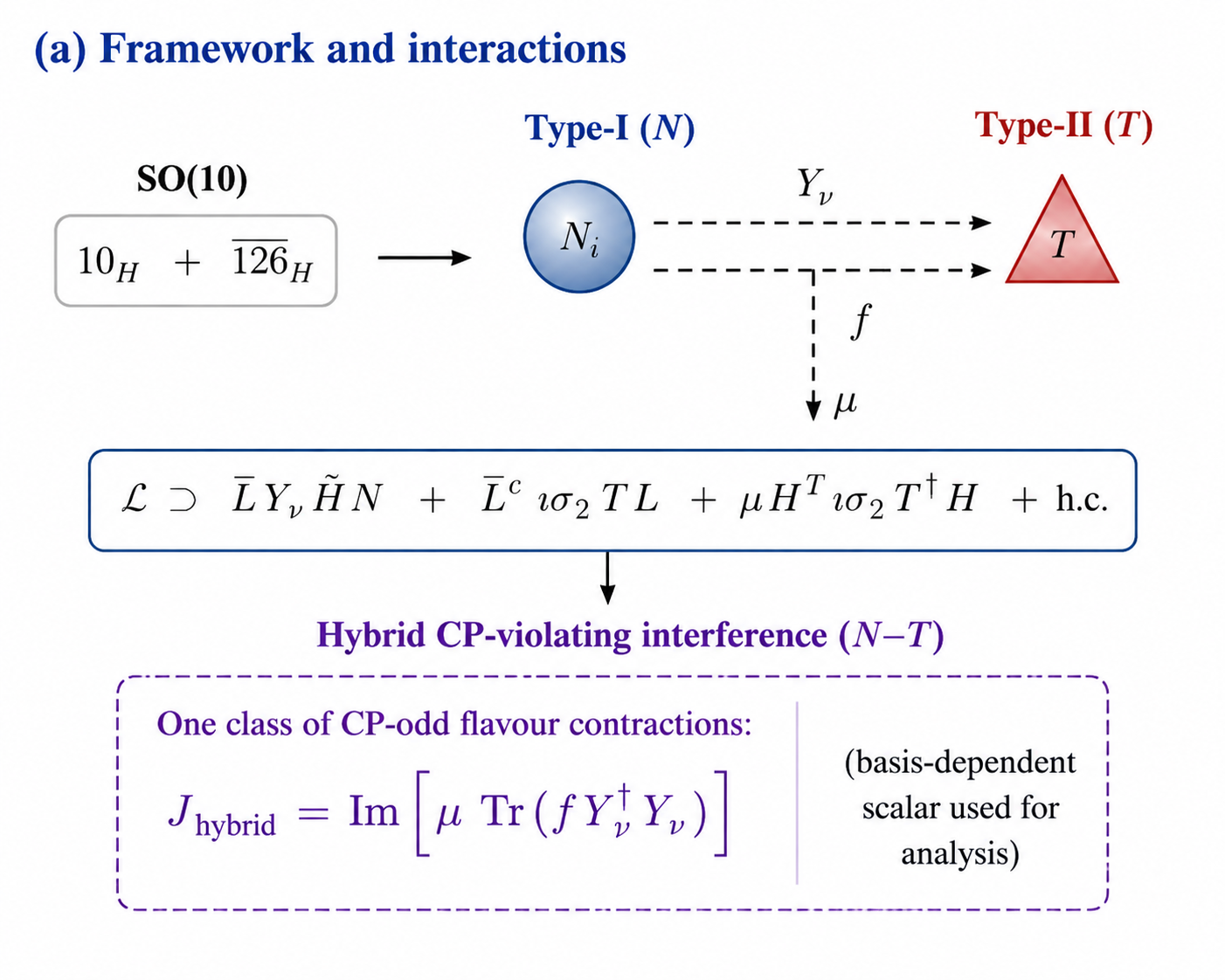}
\caption{
Schematic overview of the minimal renormalizable $\mathrm{SO}(10)$ framework considered in this work. The Yukawa interactions involving the $10_H$ and $\overline{126}_H$ Higgs representations generate both Type-I and Type-II seesaw contributions to neutrino masses. The simultaneous presence of the heavy-neutrino and scalar-triplet sectors gives rise to hybrid interference contributions to the CP asymmetry. The scalar quantity $J_{\rm hybrid}$ is introduced as a compact parametrization of one class of flavour contractions appearing in these interference amplitudes.
}
\label{fig:framework}
\end{figure}

The renormalizable Yukawa Lagrangian is given by
\begin{align}
\mathcal{L}_Y
=
16_F^T C
\left(
Y_{10}\,10_H
+
Y_{126}\,\overline{126}_H
\right)
16_F
+\mathrm{h.c.},
\end{align}
where $Y_{10}$ and $Y_{126}$ are complex symmetric matrices in flavour
space. Following symmetry breaking, the fermion mass matrices are
expressed as linear combinations of two basic structures,
\[
H \equiv v_{10}Y_{10},
\qquad
F \equiv v_{126}Y_{126},
\]
where $v_{10}$ and $v_{126}$ denote the effective electroweak vacuum
expectation values associated with the corresponding Higgs multiplets.

The resulting fermion mass matrices may be written schematically as
\begin{align}
M_u &= H + F, \\
M_d &= r_d\,H + s_d\,F, \\
M_e &= r_d\,H - 3 s_d\,F, \\
M_D &= H - 3F,
\end{align}
where the coefficients $r_d$ and $s_d$ parameterize the effects of Higgs
doublet mixing. These expressions capture the leading structure of the
fermion mass matrices in minimal renormalizable $\mathrm{SO}(10)$.
Their precise numerical realization depends on the details of the Higgs
sector, doublet mixing, threshold corrections, and renormalization-group
running between the grand-unification and electroweak scales
\cite{BajcSenjanovic,GohMohapatraNg,GrimusKuhbock,DiLuzio2021,Babu2020SO10Fit}.
Accordingly, the above relations should be regarded as the effective
low-energy structure employed in the present analysis rather than as
exact predictions of the model.

Since the same Yukawa structures determine both the fermion mass matrices
and the interactions of the heavy states, the framework naturally
provides a common origin for neutrino mass generation and leptogenesis.
The corresponding parameters may also contribute to low-energy flavour
and CP-violating observables, including charged-lepton flavour violation
and electric dipole moments. The quantitative relation between these
observables and the baryon asymmetry, however, is model dependent and
depends on the choice of Yukawa couplings, heavy-particle spectrum, and
the region of parameter space under consideration. In the following
sections we focus on the implications of this framework for hybrid
leptogenesis, where both the heavy-neutrino and scalar-triplet sectors
participate in the generation of the lepton asymmetry.
\subsection{Right-handed neutrino sector}

The $\overline{126}_H$ multiplet contains a Standard Model singlet whose
vacuum expectation value generates Majorana masses for the right-handed
neutrinos,
\begin{align}
M_R
=
v_R\,Y_{126}
=
\frac{v_R}{v_{126}}\,F
\equiv
c_R F ,
\end{align}
where $v_R$ denotes the vacuum expectation value responsible for
$B-L$ symmetry breaking, $F\equiv v_{126}Y_{126}$ is the Yukawa
structure introduced in the previous subsection, and
$c_R\equiv v_R/v_{126}$ depends on the details of the symmetry-breaking
pattern.

Depending on the Yukawa structure and the symmetry-breaking parameters,
the heavy-neutrino spectrum may be either hierarchical or
quasi-degenerate. In the present work we consider a representative
parameter region in which two heavy Majorana neutrinos satisfy
\begin{align}
|M_{N_i}-M_{N_j}|
\sim
\frac12(\Gamma_i+\Gamma_j),
\end{align}
where $\Gamma_i$ denotes the total decay width of the heavy neutrino
$N_i$. This corresponds to the resonant regime in which self-energy
corrections enhance the CP asymmetry within the heavy-neutrino sector
\cite{PilaftsisResonant,DevReviewResonant,BlanchetReview}. We emphasize
that such resonant enhancement occurs only among particles carrying
identical quantum numbers within the same sector.

\subsection{Scalar triplet sector}

The electroweak scalar triplet contained in the
$\overline{126}_H$ representation may be written as
\begin{align}
T=
\begin{pmatrix}
T^+/\sqrt{2} & T^{++}\\
T^0 & -T^+/\sqrt{2}
\end{pmatrix},
\end{align}
which transforms as $(1,3,1)$ under the Standard Model gauge group.
The normalization is chosen such that
\[
\mathrm{Tr}(T^\dagger T)
=
|T^{++}|^2+|T^+|^2+|T^0|^2 .
\]

Its renormalizable interactions with the lepton and Higgs doublets are
described by
\begin{align}
\mathcal{L}_T
\supset
\frac12
f_{\alpha\beta}\,
\overline{\ell_\alpha^c}\,
i\sigma_2
T
\ell_\beta
+
\mu
H^T
i\sigma_2
T^\dagger
H
+\mathrm{h.c.},
\end{align}
where $f_{\alpha\beta}$ is a complex symmetric Yukawa matrix. Following
electroweak symmetry breaking, the neutral component of the triplet
acquires an induced vacuum expectation value
\begin{align}
\langle T^0\rangle
\simeq
\frac{\mu v^2}{M_T^2+\lambda v^2},
\end{align}
which reduces to
\(
\langle T^0\rangle\simeq\mu v^2/M_T^2
\)
for $M_T\gg v$. The resulting Type-II seesaw contribution to the light
neutrino mass matrix is
\begin{align}
m_\nu^{\rm II}
=
2f\,\langle T^0\rangle .
\end{align}

The scalar-triplet mass is determined by the parameters of the scalar
potential and is, in general, independent of the heavy-neutrino mass
spectrum. Consequently, minimal renormalizable $\mathrm{SO}(10)$ does
not predict a quasi-degenerate relation between $M_T$ and the
right-handed neutrino masses. Nevertheless, depending on the choice of
the scalar potential and Yukawa parameters, regions of parameter space
may arise in which the scalar-triplet mass is numerically close to that
of one of the heavy Majorana neutrinos.

In the present work we consider such a parameter region as a
phenomenologically motivated benchmark. This assumption should not be
interpreted as implying resonant mixing between particles belonging to
different sectors. Rather, it provides a framework for investigating the
interplay between the Type-I and Type-II contributions to leptogenesis.
When quasi-degenerate states are present, resonant enhancement is
associated with self-energy effects occurring within the heavy-neutrino
sector or within the scalar-triplet sector separately, whereas the
hybrid contributions discussed below originate from loop-induced
interference involving both sectors.

Radiative corrections generally induce shifts in the heavy masses whose
magnitude depends on the underlying couplings. In the absence of an
additional symmetry enforcing exact mass relations, the proximity
between $M_T$ and the heavy-neutrino masses should therefore be regarded
as a phenomenological parameter choice rather than a symmetry-protected
prediction of the model.
\subsection{Hybrid neutrino masses}

The light-neutrino mass matrix receives contributions from both the
Type-I and Type-II seesaw mechanisms. Working in the flavour basis in
which the charged-lepton mass matrix and the heavy Majorana mass matrix
$M_R$ are diagonal, the effective neutrino mass matrix is given by
\begin{align}
m_\nu
=
-
M_D
M_R^{-1}
M_D^T
+
2f\langle T^0\rangle ,
\end{align}
which is valid below the heavy thresholds,
$M_R,\,M_T\gg v$. The first term corresponds to the conventional
Type-I seesaw contribution arising from the exchange of heavy
right-handed neutrinos, whereas the second term originates from the
vacuum expectation value of the electroweak scalar triplet and
represents the Type-II seesaw contribution.

The relative importance of the two contributions depends on the Yukawa
couplings and the heavy-mass spectrum. In the present work we consider a
representative region of parameter space in which the Type-II term
provides the dominant contribution to the light-neutrino mass matrix,
while the Type-I contribution remains phenomenologically relevant
through its role in the leptogenesis dynamics. This choice constitutes a
benchmark realization of the hybrid seesaw framework rather than a
generic prediction of minimal renormalizable $\mathrm{SO}(10)$.

The simultaneous presence of both seesaw contributions provides the
framework for the hybrid leptogenesis scenario studied in this work.
Their interplay influences the flavour structure entering the
CP-violating amplitudes and motivates the combined treatment of the
heavy-neutrino and scalar-triplet sectors developed in the following
sections.
\subsection{Benchmark quasi-degenerate spectrum}
In minimal renormalizable $\mathrm{SO}(10)$ models, both the
right-handed neutrino masses and the scalar-triplet mass originate from
the symmetry-breaking sector associated with the
$\overline{126}_H$ representation. The Majorana masses of the
right-handed neutrinos are generated by the vacuum expectation value of
the Standard Model singlet contained in $\overline{126}_H$, whereas the
scalar-triplet mass is determined by the parameters of the scalar
potential involving the same multiplet. Consequently, both mass scales
are associated with the underlying $B-L$ breaking scale, although their
precise values depend on independent combinations of Yukawa couplings
and scalar-potential parameters.

Because the heavy-neutrino and scalar-triplet masses are controlled by
different parameters, minimal renormalizable $\mathrm{SO}(10)$ does not
predict an exact relation between $M_T$ and $M_{N_i}$. Depending on the
choice of model parameters, however, regions of parameter space may
arise in which the scalar-triplet mass is numerically close to one of
the heavy-neutrino masses. Throughout this work we consider such a
region as a phenomenologically motivated benchmark rather than as a
generic prediction of the underlying theory.

The proximity of the two mass scales may be characterized by
\begin{equation}
\delta
=
\frac{M_T-M_3}{M_3},
\end{equation}
which measures the relative mass splitting between the scalar triplet
and the heaviest right-handed neutrino. In the numerical analysis
presented below we explore representative benchmark points with
$|\delta|\sim10^{-3}$--$10^{-5}$ in order to investigate the interplay
between the Type-I and Type-II sectors. These values are adopted as part
of the benchmark scenario considered in this work and are not intended
to represent a unique prediction of minimal renormalizable
$\mathrm{SO}(10)$.

Radiative corrections generally induce shifts in both the heavy-neutrino
and scalar-triplet masses, with magnitudes determined by the underlying
couplings and the scalar potential. In the absence of an additional
symmetry relating the two sectors, such corrections do not enforce exact
mass alignment. The quasi-degenerate regime should therefore be viewed
as a controlled choice of parameters within the model rather than as a
symmetry-protected feature.

As discussed in the following sections, this benchmark configuration
provides a useful framework for studying the interplay between the
heavy-neutrino and scalar-triplet sectors. Any resonant enhancement,
when present, arises from quasi-degenerate states within a single
sector, whereas the hybrid contributions originate from loop-induced
interference involving both sectors.
\subsection{Consistency with neutrino oscillation data}

The effective light-neutrino mass matrix receives contributions from both
the Type-I and Type-II seesaw mechanisms,
\begin{equation}
m_\nu
=
-
M_D M_R^{-1}M_D^T
+
2f\langle T^0\rangle .
\end{equation}
The relative importance of the two terms depends on the Yukawa
couplings and the heavy-particle spectrum. Throughout this work we adopt
a representative benchmark in which the Type-II contribution provides
the dominant part of the light-neutrino mass matrix, while the Type-I
term remains subleading in the neutrino masses but plays an important
role in the leptogenesis dynamics.

To ensure consistency with neutrino oscillation data, the dominant
Type-II contribution is required to reproduce the observed neutrino
mass matrix,
\begin{equation}
m_\nu
=
U_{\mathrm{PMNS}}^{*}
\,m_\nu^{\mathrm{diag}}\,
U_{\mathrm{PMNS}}^{\dagger},
\end{equation}
where
\begin{equation}
m_\nu^{\mathrm{diag}}
=
\mathrm{diag}(m_1,m_2,m_3),
\end{equation}
and $U_{\mathrm{PMNS}}$ denotes the leptonic mixing matrix.

Throughout the numerical analysis we assume normal mass ordering and
adopt the experimentally determined neutrino mass-squared differences,
\begin{align}
\Delta m_{21}^{2}
&\simeq
7.4\times10^{-5}\,\mathrm{eV}^{2},
\\
\Delta m_{31}^{2}
&\simeq
2.5\times10^{-3}\,\mathrm{eV}^{2},
\end{align}
together with leptonic mixing angles consistent with current global
fits.

Within this parametrization, the Yukawa matrix $f$ is obtained from the
relation
\begin{equation}
m_\nu^{\mathrm{II}}
=
2f\langle T^0\rangle ,
\end{equation}
so that the dominant Type-II contribution reproduces the observed
neutrino masses and mixing pattern.

The Type-I contribution,
\begin{equation}
m_\nu^{\mathrm{I}}
=
-
M_D M_R^{-1}M_D^T,
\end{equation}
is treated as a perturbation to the dominant Type-II structure. In the
benchmark scenarios considered here, the parameters are chosen such that
this correction remains consistent with the observed oscillation data
while allowing the heavy-neutrino sector to contribute to leptogenesis.

The CP-violating parameters relevant for leptogenesis are not uniquely
determined by low-energy neutrino data and are therefore treated as free
parameters within the ranges explored in the numerical analysis.

Detailed global analyses of fermion masses and mixings in minimal
renormalizable $\mathrm{SO}(10)$ have been presented in the literature
\cite{Esteban:2020cvm}. The benchmark parameter choices adopted in the
present work are chosen to be consistent with these studies while
providing a suitable framework for investigating hybrid leptogenesis.

\subsection{CP violation in the hybrid framework}

The simultaneous presence of Type-I and Type-II seesaw interactions
introduces additional sources of CP violation arising from the interplay
between the heavy-neutrino and scalar-triplet sectors. In leptogenesis,
the CP asymmetry originates from the interference between tree-level
decay amplitudes and the absorptive parts of the corresponding one-loop
corrections.

Within the present framework, hybrid contributions arise from loop
diagrams in which particles belonging to one sector appear as virtual
states in processes associated with the other sector, for example
through scalar-triplet corrections to heavy-neutrino decays and
heavy-neutrino corrections to scalar-triplet decays. These interference
terms involve both the neutrino Yukawa matrix $Y_\nu$ and the
triplet Yukawa matrix $f$, and therefore vanish if either interaction
is absent.

It is important to distinguish these hybrid contributions from the
resonant enhancement occurring in conventional resonant leptogenesis.
Since the heavy Majorana neutrinos and scalar triplets carry different
gauge quantum numbers, no Breit--Wigner resonant mixing occurs between
the two sectors. Consequently, any resonant enhancement is associated
only with quasi-degenerate particles within the heavy-neutrino sector
or within the scalar-triplet sector separately, whereas the hybrid
contributions constitute additional loop-induced corrections to the
CP asymmetry.

\subsection{Scalar flavour contraction for the hybrid interference terms}

To facilitate the discussion of the flavour structure entering the
hybrid interference amplitudes, it is convenient to introduce the scalar
combination

\begin{equation}
J_{\rm hybrid}
=
\mathrm{Im}
\!\left[
\mu\,
\mathrm{Tr}
\!\left(
fY_\nu^\dagger Y_\nu
\right)
\right],
\label{eq:Jhybrid}
\end{equation}

where $\mu$ denotes the trilinear scalar coupling appearing in the
scalar potential and $f$ and $Y_\nu$ are the triplet and neutrino Yukawa
coupling matrices, respectively.

The quantity defined in Eq.~(\ref{eq:Jhybrid}) provides a compact scalar
parametrization of one class of flavour contractions entering the hybrid
interference amplitudes within the flavour basis adopted throughout this
work. It vanishes when either $Y_\nu$ or $f$ is absent, reflecting the
fact that the corresponding interference contributions require the
simultaneous presence of both Type-I and Type-II interactions.

The quantity $J_{\rm hybrid}$ is introduced as a convenient measure of
the flavour structure appearing in the hybrid amplitudes considered in
the present analysis. We do not interpret it as providing a complete
characterization of CP violation in the general theory, since additional
CP-odd flavour structures may contribute depending on the choice of
basis and the underlying parameter space.

\section{Hybrid CP asymmetries}
\begin{figure}[t]
\centering
\includegraphics[width=0.42\textwidth]{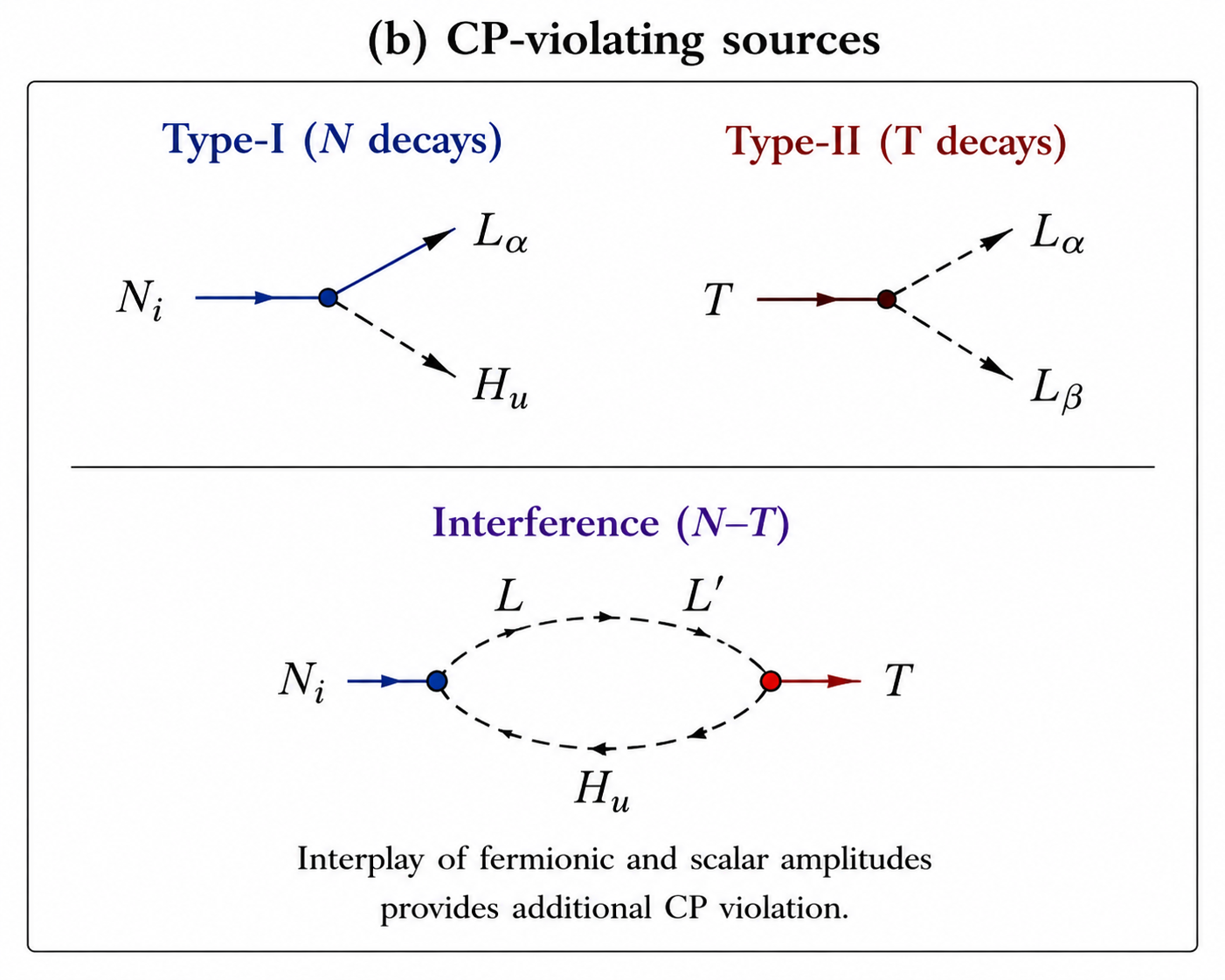}
\caption{
Schematic illustration of the CP-violating sources in the hybrid
Type-I+Type-II leptogenesis framework. The upper panels show the
conventional CP asymmetries generated in heavy-neutrino ($N_i$) and
scalar-triplet ($T$) decays. The lower panel illustrates the additional
hybrid interference between the fermionic and scalar sectors, which
generates CP-violating contributions only when both sectors are
simultaneously present. Resonant enhancement, when applicable, occurs
only among quasi-degenerate states within the same sector, whereas the
hybrid interference terms discussed here remain non-resonant.
}
\label{fig:CPsources}
\end{figure}

In the framework considered in this work, both Type-I and Type-II
seesaw interactions contribute to the generation of the lepton
asymmetry. In addition to the conventional CP asymmetries arising
independently from heavy-neutrino and scalar-triplet decays,
loop-induced interference between the fermionic and scalar sectors
gives rise to additional hybrid contributions. These contributions
are present only when both interaction sectors are simultaneously
active and therefore vanish in the limits
\(
Y_\nu = 0
\)
or
\(
f = 0,
\)
where the dynamics reduces to the conventional Type-I or Type-II
leptogenesis scenarios. Figure~\ref{fig:CPsources} provides a
schematic overview of the different CP-violating sources considered
in the present framework and illustrates the interplay between the
fermionic and scalar sectors responsible for the hybrid interference
terms discussed below.

As in conventional leptogenesis, CP asymmetries originate from the
interference between tree-level decay amplitudes and the absorptive
parts of the corresponding one-loop corrections. In the present
framework, additional loop diagrams involving virtual particles from the
complementary sector contribute to the decay amplitudes of both heavy
Majorana neutrinos and scalar triplets. These interference terms depend
on the flavour structure of both the neutrino Yukawa couplings and the
triplet Yukawa interactions, thereby providing additional contributions
to the CP asymmetry beyond those arising within either sector
individually.
Throughout this section we distinguish carefully between resonant
enhancement and hybrid interference effects. Resonant enhancement, when
present, is associated with quasi-degenerate states carrying identical
quantum numbers within a single sector and is treated using the standard
resummed propagator formalism. In contrast, the hybrid contributions
discussed below originate from loop-induced interference between the
heavy-neutrino and scalar-triplet sectors and do not correspond to
Breit--Wigner resonant mixing between particles with different quantum
numbers. This distinction will be maintained throughout the analytical
and numerical discussions that follow.
\subsection{Origin of hybrid CP asymmetries}

The hybrid contributions considered in this work originate from the
interplay between the heavy-neutrino and scalar-triplet sectors. The
relevant decay processes include
\begin{itemize}
\item $N_i \rightarrow \ell_\alpha H$, with one-loop corrections
containing virtual scalar triplets,
\item $T \rightarrow \ell_\alpha \ell_\beta$, with one-loop corrections
containing virtual heavy Majorana neutrinos.
\end{itemize}

The corresponding CP asymmetries arise from the interference between the
tree-level amplitudes and the absorptive parts of the associated
one-loop diagrams. Since these loop contributions involve both the
neutrino Yukawa couplings $Y_\nu$ and the triplet Yukawa couplings $f$,
they vanish when either interaction is absent. They therefore constitute
additional interference contributions beyond the conventional Type-I and
Type-II leptogenesis mechanisms.

It is important to distinguish these hybrid interference effects from
the resonant enhancement encountered in conventional resonant
leptogenesis. Resonant enhancement occurs only among quasi-degenerate
particles carrying identical quantum numbers within a given sector. For
the heavy-neutrino sector this corresponds to
\begin{align}
|M_{N_i}-M_{N_j}|
\sim
\frac{1}{2}
(\Gamma_i+\Gamma_j),
\end{align}
where $\Gamma_i$ denotes the total decay width of the heavy neutrino
$N_i$. A similar condition may be realized in models containing
multiple quasi-degenerate scalar triplets. By contrast, no resonant
enhancement occurs between particles belonging to different sectors,
such as heavy Majorana neutrinos and scalar triplets, since these states
carry different gauge quantum numbers and do not undergo resonant
mixing.

\subsection{Treatment of resonant enhancement and regulator structure}

The CP asymmetries discussed in the following subsections contain
self-energy contributions that require special treatment in the
quasi-degenerate regime. In order to avoid ambiguities associated with
unstable intermediate states, we briefly summarize the theoretical
framework adopted in the present analysis.

\subsubsection*{Formalism}

The resonant contributions are evaluated using the standard resummed
propagator formalism commonly employed in resonant leptogenesis. Within
this approach, the enhancement of the CP asymmetry originates from the
interference between tree-level amplitudes and one-loop self-energy
corrections, while finite-width effects are consistently incorporated
through resummed unstable propagators.

The resummed treatment provides a well-established approximation for
computing CP asymmetries in the quasi-degenerate regime and is
consistent with the leading-order description obtained from
non-equilibrium quantum field theory in the appropriate limit. It has
been widely applied in studies of resonant leptogenesis and provides the
framework adopted throughout the present work.

\subsubsection*{Regulator and decay widths}

For quasi-degenerate states within a given sector, the self-energy
contribution to the CP asymmetry contains the familiar resonant
regulator
\begin{equation}
\frac{\Delta M_{ij} M_i \Gamma_j}
{(\Delta M_{ij})^2 + (M_i \Gamma_j)^2},
\end{equation}
where
$\Delta M_{ij}=M_i-M_j$
denotes the mass splitting between the intermediate states and
$\Gamma_j$ is the corresponding total decay width.

Throughout this work, the decay widths entering the regulator are
evaluated at tree level. For heavy Majorana neutrinos we use
\begin{equation}
\Gamma_i
=
\frac{(Y_\nu^\dagger Y_\nu)_{ii}}
{8\pi}
M_i,
\end{equation}
while for scalar triplets the total width is taken as
\begin{equation}
\Gamma_T
=
\Gamma(T\rightarrow\ell\ell)
+
\Gamma(T\rightarrow HH),
\end{equation}
as defined in Eq.~(17).

These expressions are employed consistently throughout the numerical
analysis and are appropriate within the narrow-width approximation,
where the finite lifetime of the unstable intermediate states is
incorporated through the regulator above.

\subsubsection*{Avoidance of double counting}

A well-known issue in resonant leptogenesis is the potential double
counting between decay and scattering processes when intermediate
particles can become on shell. In the present analysis, the CP
asymmetries are computed at the level of decay amplitudes using resummed
propagators, whereas washout effects are incorporated separately through
inverse decays and scattering processes in the Boltzmann equations.

This procedure avoids double counting of on-shell intermediate states
and is consistent with the real intermediate state (RIS) subtraction
prescription commonly employed in the literature.

\subsubsection*{Scope and limitations}

The regulator introduced above applies exclusively to resonant
self-energy contributions arising from quasi-degenerate particles within
the same sector. For example, in the heavy-neutrino sector resonant
enhancement occurs only for quasi-degenerate right-handed neutrinos,
while an analogous treatment applies to models containing multiple
quasi-degenerate scalar triplets.

The hybrid contributions investigated in the present work involve both
heavy Majorana neutrinos and scalar triplets. Since these particles
carry different gauge quantum numbers and do not undergo resonant
mixing, the corresponding loop contributions do not exhibit
Breit--Wigner resonant enhancement. Instead, they represent additional
loop-induced interference effects whose behaviour is determined by the
corresponding threshold structure of the loop functions discussed in
Sec.~II.D.

The treatment adopted throughout this work therefore combines the
standard resonant formalism for quasi-degenerate states within a given
sector with the non-resonant hybrid interference contributions arising
from the simultaneous presence of the Type-I and Type-II interactions.
\subsection{Heavy-neutrino decays}

The flavoured CP asymmetry associated with the decay
$N_i\rightarrow\ell_\alpha H$ is defined by
\begin{align}
\varepsilon_{i\alpha}
=
\frac{
\Gamma(N_i\rightarrow\ell_\alpha H)
-
\Gamma(N_i\rightarrow\bar{\ell}_\alpha H^\dagger)
}{
\displaystyle
\sum_\beta
\left[
\Gamma(N_i\rightarrow\ell_\beta H)
+
\Gamma(N_i\rightarrow\bar{\ell}_\beta H^\dagger)
\right]
}.
\end{align}

Within the hybrid framework, the CP asymmetry may be written as the sum
of the conventional Type-I contribution and an additional interference
term,
\begin{align}
\varepsilon_{i\alpha}
=
\varepsilon^{\rm I}_{i\alpha}
+
\varepsilon^{\rm hybrid}_{i\alpha},
\end{align}
where $\varepsilon^{\rm I}_{i\alpha}$ denotes the standard Type-I
leptogenesis contribution.

The hybrid contribution originates from one-loop diagrams containing
virtual scalar-triplet states. A representative contribution may be
written schematically as
\begin{align}
\varepsilon^{\rm hybrid}_{i\alpha}
&\sim
\frac{1}
{8\pi (Y_\nu Y_\nu^\dagger)_{ii}}
\sum_{j\neq i}
\operatorname{Im}
\!\left[
(Y_\nu^\dagger fY_\nu^*)_{ij}
(Y_\nu Y_\nu^\dagger)_{ji}
\right]
\nonumber\\
&\qquad\times
\mathcal{F}_T
\!\left(
M_i^2,
M_j^2,
M_T^2
\right),
\end{align}
where $\mathcal{F}_T$ denotes the loop function arising from the
corresponding vertex and self-energy corrections. The above expression
illustrates the flavour structure entering the hybrid interference
terms; the complete CP asymmetry receives contributions from the full
set of one-loop amplitudes considered in the present analysis.

The hybrid contribution vanishes if either the neutrino Yukawa coupling
$Y_\nu$ or the triplet Yukawa coupling $f$ is absent, reflecting the
fact that both interaction sectors are required for the corresponding
interference effects.

For quasi-degenerate heavy neutrinos, the conventional self-energy
contribution within the heavy-neutrino sector may be resonantly
enhanced and therefore requires the standard resummed propagator
treatment\cite{Pilaftsis1999}.
By contrast, the loop function
$\mathcal{F}_T(M_i^2,M_j^2,M_T^2)$
describes the threshold behaviour of the hybrid interference
contribution. Since the virtual scalar triplet and the heavy Majorana
neutrinos belong to different sectors and do not undergo resonant
mixing, no Breit--Wigner regulator is introduced for this part of the
amplitude.
\subsection{Scalar-triplet decays}

The CP asymmetry associated with scalar-triplet decays is defined by
\begin{align}
\varepsilon_T
=
\frac{
\Gamma(T\rightarrow\ell\ell)
-
\Gamma(T\rightarrow\bar{\ell}\bar{\ell})
}{
\Gamma_T
},
\qquad
\Gamma_T
=
\Gamma(T\rightarrow\ell\ell)
+
\Gamma(T\rightarrow HH),
\end{align}
where $\Gamma_T$ denotes the total decay width of the scalar triplet.

In addition to the conventional Type-II contribution, the hybrid
framework considered here receives interference contributions from
one-loop diagrams containing virtual heavy Majorana neutrinos. A
representative contribution may be written schematically as
\begin{align}
\varepsilon_T^{\rm hybrid}
&\sim
\frac{1}{8\pi\,\Gamma_T}
\sum_i
\operatorname{Im}
\!\left[
\mu\,
\mathrm{Tr}
\!\left(
fY_\nu^\dagger Y_\nu
\right)
\right]
\nonumber\\
&\qquad\times
\mathcal{G}_N
\!\left(
M_T^2,
M_i^2
\right),
\end{align}
where $\mathcal{G}_N$ denotes the loop function associated with virtual
heavy-neutrino exchange. The expression above illustrates one
representative flavour contraction appearing in the hybrid interference
terms, while the complete CP asymmetry receives contributions from the
full set of one-loop amplitudes discussed in the present analysis.

The hybrid contribution vanishes in the absence of either the triplet
Yukawa coupling $f$ or the neutrino Yukawa coupling $Y_\nu$,
demonstrating that both interaction sectors are required for the
corresponding interference effects.

Unlike the conventional resonant enhancement occurring for
quasi-degenerate particles within a single sector, the hybrid
contribution does not exhibit Breit--Wigner resonant behaviour. Since
the scalar triplet and the heavy Majorana neutrinos carry different
gauge quantum numbers and do not mix, the loop function
$\mathcal{G}_N(M_T^2,M_i^2)$ describes the threshold dependence of the
hybrid interference contribution rather than resonant propagation.
\subsection{Kinetic evolution, baryon asymmetry and low-energy observables}

The generation of the cosmological baryon asymmetry results from the
combined evolution of the heavy-neutrino and scalar-triplet sectors,
including CP-violating decays, inverse decays, scattering processes and
flavour oscillations. In the present work the evolution of the lepton
asymmetry is described within the flavour-covariant density-matrix
formalism, which consistently incorporates flavour coherence and
washout effects.

The comoving lepton asymmetry is represented by the flavour-space matrix
$Y_{\Delta\ell}$ satisfying
\begin{equation}
\frac{dY_{\Delta\ell}}{dz}
=
S(z)
-
\frac12
\left\{
\Gamma(z),
Y_{\Delta\ell}
\right\}
-
i
\left[
H(z),
Y_{\Delta\ell}
\right],
\end{equation}
where
$z=M/T$
denotes the dimensionless inverse temperature.

The source term $S(z)$ contains the CP asymmetries generated in both
heavy-neutrino and scalar-triplet decays, including the hybrid
interference contributions discussed above. The matrix
$\Gamma(z)$ describes washout effects arising from inverse decays and
scattering processes, while
$H(z)$ denotes the effective Hamiltonian governing flavour oscillations
induced by charged-lepton Yukawa interactions.

The abundances of the heavy particles evolve according to
\begin{align}
\frac{dY_{N_i}}{dz}
&=
-
D_i(z)
\left(
Y_{N_i}
-
Y_{N_i}^{\rm eq}
\right),
\\
\frac{dY_T}{dz}
&=
-
D_T(z)
\left(
Y_T
-
Y_T^{\rm eq}
\right),
\end{align}
where $D_i(z)$ and $D_T(z)$ denote the corresponding decay terms.

Within the flavour basis adopted throughout this work, the hybrid
interference amplitudes involve flavour contractions containing both the
neutrino Yukawa matrix $Y_\nu$ and the triplet Yukawa matrix $f$. The
scalar quantity
$J_{\rm hybrid}$ introduced in
Eq.~(\ref{eq:Jhybrid})
provides a convenient parametrization of one representative flavour
contraction appearing in these amplitudes. The complete CP asymmetries,
however, are obtained from the full set of tree-level and one-loop
contributions discussed in the preceding subsections.

After solving the kinetic equations, the baryon asymmetry is obtained
through electroweak sphaleron conversion,
\begin{equation}
Y_B
=
c_{\rm sph}
\,\mathrm{Tr}
(Y_{\Delta\ell}),
\qquad
c_{\rm sph}
=
\frac{28}{79},
\end{equation}
where the trace is taken over flavour indices.

For qualitative guidance, the final baryon asymmetry may be represented
schematically as
\begin{align}
Y_B
\propto
\sum_i
\kappa_i
\,
\varepsilon_i
+
\kappa_T
\,
\varepsilon_T,
\end{align}
where the efficiency factors
$\kappa_i$
and
$\kappa_T$
summarize the effects of washout, inverse decays, scattering processes
and spectator interactions.

The same Yukawa interactions responsible for leptogenesis may also
generate charged-lepton flavour violation and CP-violating observables
at low energies. To illustrate the parametric dependence on the model
parameters, the corresponding amplitudes may be represented
schematically by
\begin{align}
{\rm BR}
(\mu\rightarrow e\gamma)
&\sim
\frac{\alpha}{192\pi}
\frac{|(ff^\dagger)_{e\mu}|^2}
{M_T^4},
\\
d_e
&\sim
\frac{em_e}
{16\pi^2M_T^2}
{\rm Im}
[(ff^\dagger)_{ee}],
\end{align}
where the above expressions indicate only the leading parametric
scaling. A complete quantitative prediction requires the full one-loop
or two-loop amplitudes, including the appropriate loop functions,
electroweak couplings and numerical coefficients. Consequently, these
relations are employed only to identify the relevant regions of
parameter space rather than to provide precision predictions.

\subsubsection*{Validity of the approximations}

The kinetic equations employed above correspond to the leading-order
density-matrix description derived from non-equilibrium quantum field
theory. They are appropriate in the temperature regime where flavour
coherence and decoherence effects play an important role while higher-
order quantum corrections remain subdominant.

Throughout the numerical analysis we assume thermal initial abundances
for the heavy particles and include the dominant washout and flavour
effects at leading order. Higher-order spectator corrections and
subleading quantum effects are neglected. These approximations provide a
self-consistent framework for studying the interplay between Type-I,
Type-II and hybrid leptogenesis over the parameter space explored in the
present work.
\section{Numerical Strategy and Benchmark Choice}

\begin{figure}[t]
\centering
\includegraphics[width=0.48\textwidth]{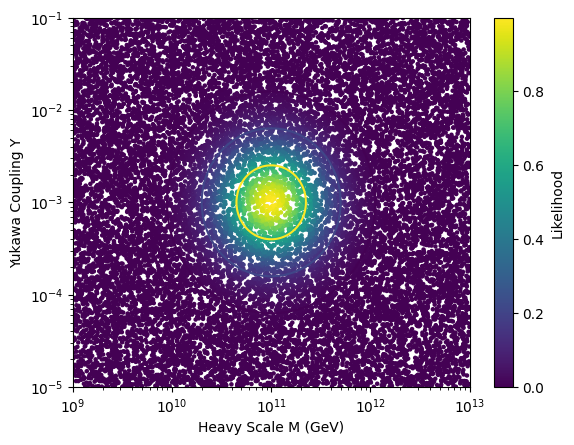}
\caption{
Numerical scan of the model parameter space in the heavy-mass--Yukawa
plane. Each point corresponds to one parameter realization generated
within the ranges
$10^{9}\le M\le10^{13}\,\mathrm{GeV}$ and
$10^{-5}\le Y\le10^{-1}$.
The colour scale indicates the corresponding value of the effective
$\Delta\chi^{2}$ constructed from the theoretical consistency
requirements and phenomenological constraints described in the text.
The white contours indicate representative regions of increasing
$\Delta\chi^{2}$, corresponding to
$\Delta\chi^{2}=2.3,\;6.18,\;11.83$.
The concentration of viable parameter points reflects the correlated
dependence of the heavy mass scales and Yukawa couplings required for
successful leptogenesis while satisfying the imposed theoretical and
phenomenological constraints. The benchmark points adopted in the
numerical analysis are selected from representative regions of the
viable parameter space.
}
\label{fig:scan}
\end{figure}

To investigate the phenomenological implications of the hybrid
leptogenesis framework, we perform a numerical exploration of the model
parameter space by varying the heavy-particle masses, Yukawa couplings,
and CP-violating parameters over broad perturbative ranges. The
resulting distribution of viable parameter points is shown in
Fig.~\ref{fig:scan}. Each point corresponds to a complete realization
of the model, while the colour scale represents the corresponding value
of $\Delta\chi^2$ obtained from the theoretical consistency conditions
and phenomenological constraints discussed in the previous sections.

The numerical scan reveals localized regions of parameter space in which
the observed baryon asymmetry can be reproduced consistently with the
imposed theoretical and phenomenological requirements. The preferred
region reflects the interplay between the heavy mass scales, the Yukawa
couplings and the efficiency of washout processes. In particular,
larger Yukawa couplings generally require larger heavy-particle masses
to maintain successful leptogenesis, leading to the correlated
distribution visible in Fig.~\ref{fig:scan}. The scan therefore serves
to identify representative regions of phenomenological viability rather
than isolated parameter points.

To illustrate the characteristic behaviour of the model, we select
benchmark configurations from three representative regions of the viable
parameter space, namely the region with the smallest $\Delta\chi^2$, an
intermediate region, and the boundary of phenomenological viability.
These benchmark points are intended to typify extended regions of the
parameter space explored in the scan rather than unique parameter
choices.

Typical viable solutions are characterized by heavy mass scales
\[
M_{N,T}\sim10^{10\text{--}12}\,\mathrm{GeV},
\]
perturbative Yukawa couplings
\[
Y\sim10^{-4}\text{--}10^{-2},
\]
and CP-violating phases of order unity. Within this region, the
generated CP asymmetry and the corresponding washout effects combine to
yield a baryon asymmetry compatible with the observed cosmological
value.

For definiteness, one representative benchmark from the preferred
region of the numerical scan is adopted throughout the following
analysis. The heavy-neutrino spectrum is chosen to be hierarchical,
\[
M_1\ll M_2\ll M_3,
\]
with
\begin{align}
M_1 &= 10^{7}\,\mathrm{GeV}, \nonumber\\
M_2 &= 10^{9}\,\mathrm{GeV}, \nonumber\\
M_3 &= 1.0\times10^{11}\,\mathrm{GeV}.
\end{align}

The scalar-triplet mass is chosen to lie close to the mass of the
heaviest right-handed neutrino,
\[
M_T=M_3(1+\delta),
\qquad
|\delta|=10^{-3},
\]
providing a representative benchmark for investigating the threshold
behaviour of the hybrid loop contributions. As discussed in
Sec.~II, this quasi-degenerate configuration is adopted as a
phenomenological benchmark rather than as a generic prediction of
minimal renormalizable $\mathrm{SO}(10)$. Since the scalar triplet and
the heavy Majorana neutrinos belong to different sectors and possess
different gauge quantum numbers, this mass proximity does not imply
Breit--Wigner resonant enhancement
\cite{DevPilaftsisReview}.

The remaining benchmark parameters are chosen as
\begin{align}
|f| &\sim 10^{-2}, \nonumber\\
|\mu| &= 10^{9}\,\mathrm{GeV},
\end{align}
while the CP-violating phases are treated as free parameters within the
ranges explored in the numerical scan. These values lie within the
perturbative regime and are representative of the parameter space
considered in the present analysis.

Thermal initial abundances are assumed for both the scalar triplet and
the heavy Majorana neutrinos, while the initial lepton asymmetry is
taken to vanish. The subsequent evolution is obtained by solving the
flavour-covariant kinetic equations introduced in the previous section,
including the effects of inverse decays, scattering processes and
flavour oscillations. The resulting baryon asymmetry is obtained after
electroweak sphaleron conversion according to
\[
Y_B
=
c_{\rm sph}\,
\mathrm{Tr}(Y_{\Delta\ell}),
\qquad
c_{\rm sph}
=
\frac{28}{79},
\]
where the Standard Model sphaleron conversion coefficient is used
consistently throughout the analysis.

For the benchmark configurations considered here, the numerical
solutions reproduce the observed baryon asymmetry,
\[
Y_B\simeq8.7\times10^{-11},
\]
for representative heavy mass scales around
$10^{11}\,\mathrm{GeV}$ and perturbative Yukawa couplings. These
benchmark points therefore illustrate that successful hybrid
leptogenesis can be realized within the region of parameter space
explored in the present study.

\subsection{Parameter-space scan and benchmark selection}

The multidimensional parameter space of the model was explored through a
random scan over the heavy-neutrino masses, scalar-triplet mass, Yukawa
couplings and CP-violating parameters. Only parameter points satisfying
theoretical consistency conditions, successful baryogenesis, and current
experimental constraints from charged-lepton flavour violation and
electric dipole moment searches were retained.

\begin{figure}[t]
\centering
\includegraphics[width=0.43\textwidth]{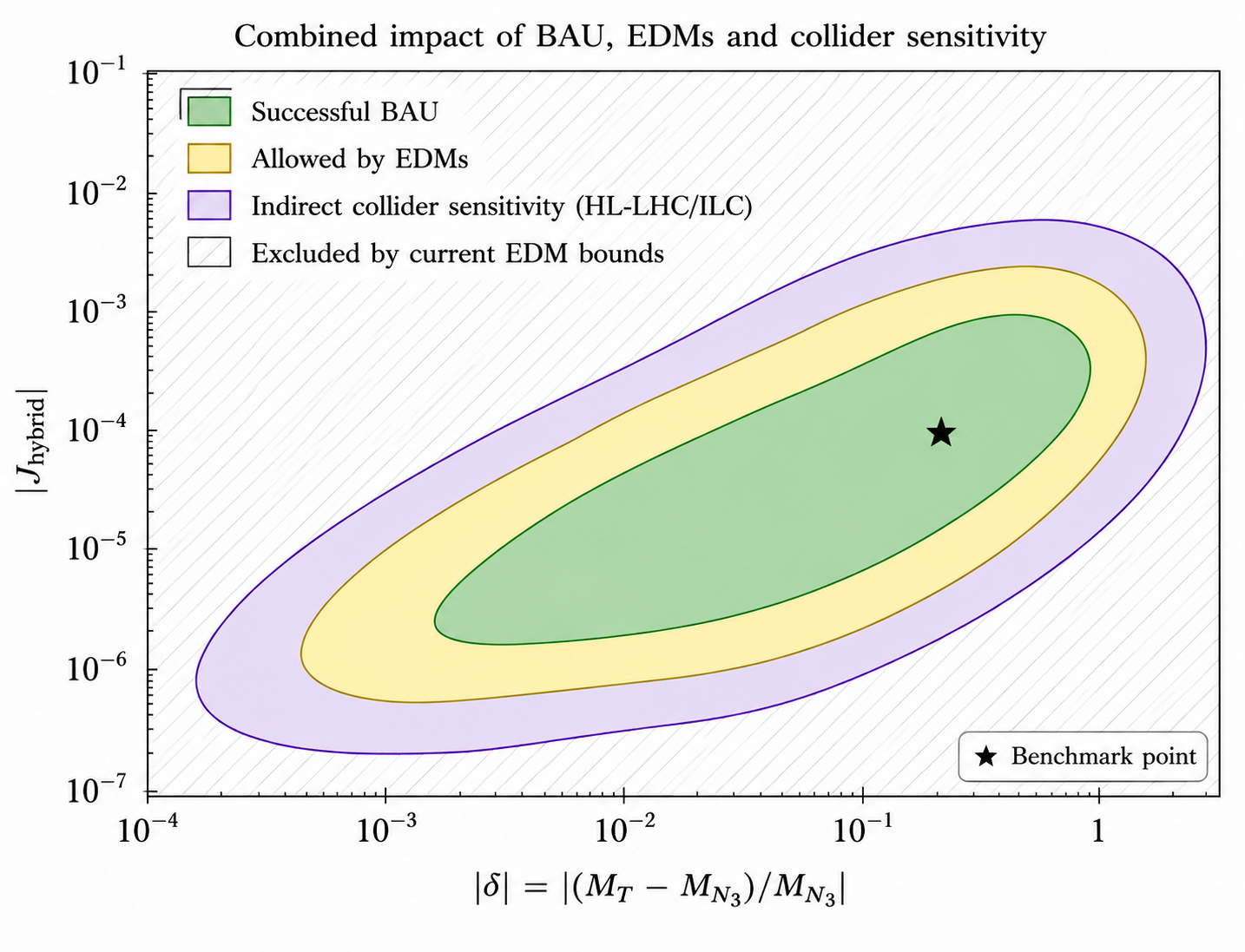}
\caption{
Allowed parameter space of the hybrid leptogenesis scenario in the plane
of the hybrid CP-violating invariant $|J_{\rm hybrid}|$ and the
quasi-degeneracy parameter
$|\delta|=|(M_T-M_{N_3})/M_{N_3}|$.
The nested regions correspond to successful baryogenesis (green),
parameter points consistent with present electric dipole moment limits
(gold), and the parameter space potentially accessible through indirect
collider probes (purple). The hatched region is excluded by current EDM
constraints. The star denotes the benchmark point used throughout the
numerical analysis.
}
\label{fig:parameterspace}
\end{figure}

Figure~\ref{fig:parameterspace} summarizes the resulting allowed
parameter space in the plane of the hybrid CP-violating invariant
$|J_{\rm hybrid}|$ and the quasi-degeneracy parameter
$|\delta|=|(M_T-M_{N_3})/M_{N_3}|$. The nested coloured regions indicate
parameter domains compatible with successful baryogenesis, present EDM
constraints, and projected indirect collider sensitivity, while the star
marks the representative benchmark point adopted in the subsequent
numerical analysis.

The allowed region forms a continuous band rather than isolated points,
indicating that successful hybrid leptogenesis does not rely on extreme
fine tuning. Instead, the viable parameter space is determined by the
interplay between CP-asymmetry generation, washout effects, and current
experimental constraints. The benchmark point indicated in
Fig.~\ref{fig:parameterspace} is selected from the central high-likelihood
region and is used in the numerical solutions presented in the following
section.
\subsection{Numerical implementation and reproducibility}

The coupled system of kinetic equations governing the evolution of the
heavy-neutrino abundances, the scalar-triplet abundance and the
flavour-dependent lepton asymmetry is solved numerically using an
adaptive step-size Runge--Kutta integration algorithm. The adaptive
integration procedure provides stable numerical solutions over the
temperature range relevant for leptogenesis, including regions where the
reaction rates vary rapidly.

The initial conditions correspond to thermal abundances of the heavy
particles,
\begin{equation}
Y_{N_i}(z\ll1)
=
Y_{N_i}^{\rm eq},
\qquad
Y_T(z\ll1)
=
Y_T^{\rm eq},
\end{equation}
together with a vanishing initial lepton asymmetry,
\begin{equation}
Y_{\Delta\ell}(z\ll1)=0.
\end{equation}
These assumptions are adopted throughout the numerical analysis and are
consistent with the benchmark scenario introduced in the previous
section.

Thermal effects are included at leading order through the
temperature-dependent equilibrium abundances and reaction rates entering
the kinetic equations. Higher-order thermal corrections, including
thermal masses and quantum-statistical effects, are not incorporated
explicitly. Throughout the benchmark region explored in this work, these
contributions are expected to provide subleading corrections to the
leading-order evolution considered here.

The numerical implementation has been tested by varying the integration
step size and confirming the convergence of the resulting baryon
asymmetry. In addition, the parameter-space scan is performed using
logarithmic priors for the heavy masses and Yukawa couplings, and the
stability of the scan has been examined by increasing the density of
randomly generated parameter points.

The numerical uncertainty associated with the integration procedure is
estimated from the convergence tests to be at the level of a few
percent, which is smaller than the theoretical uncertainties associated
with the leading-order treatment of the leptogenesis dynamics adopted in
the present analysis.
\subsection{Interpretation of the parameter scan}

The global structure of the parameter space explored in this work is
illustrated in Fig.~\ref{fig:scan}, which shows the results of the
numerical scan in the heavy-mass--Yukawa plane. Each point corresponds
to a complete parameter realization of the model, while the colour scale
represents the value of the effective $\Delta\chi^2$ constructed from
the theoretical consistency conditions and phenomenological constraints
described in the text. The scan ranges are chosen sufficiently broad to
sample the relevant regions of parameter space over the perturbative
domain considered in this analysis.

Figure~\ref{fig:scan} exhibits a correlated distribution of viable
parameter points resulting from the interplay between the heavy mass
scales, Yukawa couplings, CP-violating parameters and washout effects.
Successful leptogenesis is obtained only within restricted regions of
parameter space where the generated CP asymmetry is sufficiently large
to reproduce the observed baryon asymmetry while remaining compatible
with perturbativity and the phenomenological constraints imposed in the
analysis. The diagonal structure visible in the figure reflects the
correlation between the heavy mass scale and the Yukawa couplings,
whereby larger Yukawa couplings generally require larger heavy masses to
maintain a viable baryon asymmetry.

The white contour lines shown in Fig.~\ref{fig:scan} correspond to
representative regions of increasing effective
$\Delta\chi^2$, as defined in the numerical analysis. The innermost
contour identifies the region with the lowest effective
$\Delta\chi^2$, while the outer contours indicate progressively less
preferred regions that nevertheless satisfy the imposed theoretical and
phenomenological requirements. The continuous distribution of viable
points illustrates that successful leptogenesis is obtained throughout
an extended region of the parameter space explored in the present scan.

Parameter points outside the outermost contour generally fail to satisfy
one or more of the imposed constraints. Depending on the region of
parameter space, this may arise from an insufficient CP asymmetry,
excessive washout effects, violation of perturbativity, or
incompatibility with the phenomenological bounds considered in this
work. The numerical scan therefore illustrates how these requirements
collectively restrict the viable parameter space of the model.

To illustrate representative realizations of the hybrid leptogenesis
mechanism, benchmark configurations are selected from three regions of
the numerical scan, namely the region with the lowest effective
$\Delta\chi^2$, an intermediate region, and the boundary of the viable
parameter space. These benchmark points are intended to represent
typical solutions within the scan rather than unique parameter choices.
Representative viable configurations are characterized by heavy mass
scales
\[
M\sim10^{10\text{--}12}\,\mathrm{GeV},
\]
and perturbative Yukawa couplings
\[
Y\sim10^{-4}\text{--}10^{-2},
\]
for which the numerical solutions yield a baryon asymmetry compatible
with the observed cosmological value.

Overall, the numerical scan reveals a correlated region of parameter
space in which hybrid leptogenesis can successfully account for the
observed baryon asymmetry within the assumptions of the present
framework. Future improvements in experimental searches for
charged-lepton flavour violation and electric dipole moments may provide
additional constraints on portions of this parameter space, thereby
offering complementary tests of the underlying model.
\begin{figure}[t]
\centering
\includegraphics[width=0.4\textwidth]{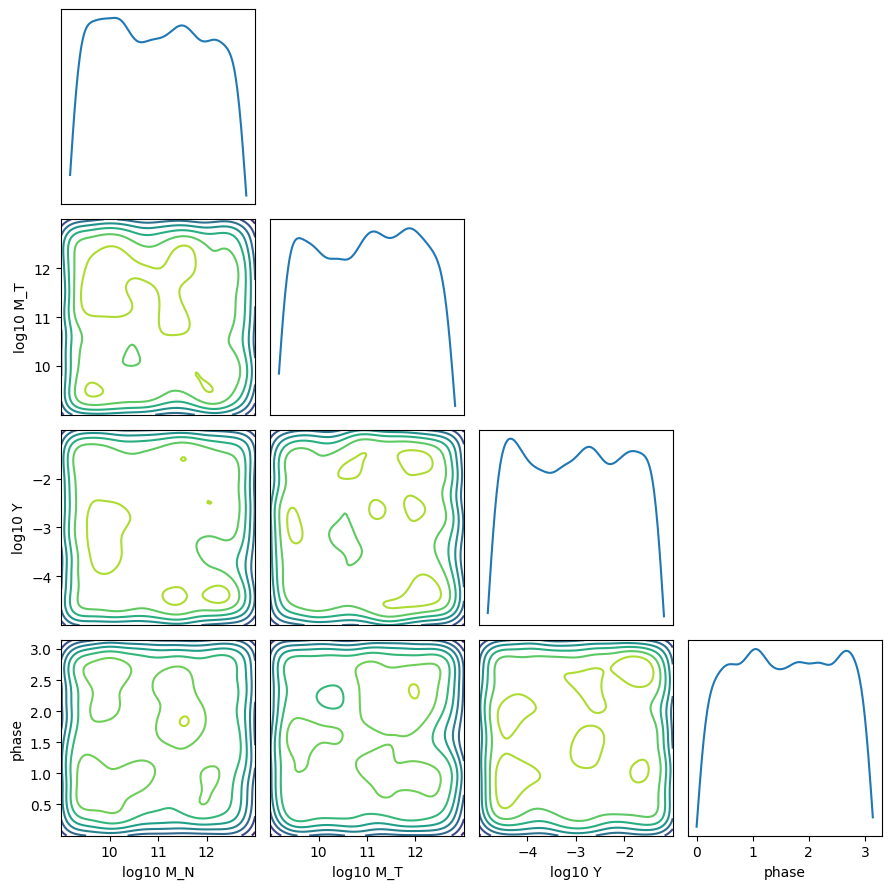}
\caption{
Corner plot illustrating the distribution of accepted parameter points
obtained from the numerical scan of the hybrid leptogenesis framework.
The scanned parameters include the heavy-neutrino mass $M_N$, the
scalar-triplet mass $M_T$, the Yukawa coupling $Y$, and the scalar
quantity $J_{\mathrm{hybrid}}$ introduced in the text. The diagonal
panels show the distributions of the accepted scan points for the
individual parameters, while the off-diagonal panels display the
corresponding two-dimensional projections, illustrating correlations
among the model parameters. Only parameter points satisfying the
theoretical consistency conditions and the phenomenological constraints
imposed in the numerical analysis are retained, including the current
bound
$\mathrm{BR}(\mu\rightarrow e\gamma)<1.5\times10^{-13}$
\cite{MEGII2024}.
}
\label{fig:corner_constraints}
\end{figure}
Figure~\ref{fig:corner_constraints} provides a global visualization of
the parameter space explored in the numerical analysis after imposing
the theoretical consistency conditions and the phenomenological
constraints discussed in the previous sections. The diagonal panels show
the distributions of the accepted scan points for each parameter,
illustrating the ranges that remain compatible with the requirements
imposed in the analysis. The off-diagonal panels display the
corresponding two-dimensional projections and reveal the correlations
among the model parameters that arise from the simultaneous requirements
of successful leptogenesis and phenomenological consistency.

The accepted parameter points satisfy the experimental constraints on
charged-lepton flavour violation and CP-violating observables. In
particular, the numerical scan incorporates the bound
\[
\mathrm{BR}(\mu\rightarrow e\gamma)
<
1.5\times10^{-13},
\]
reported by the MEG~II experiment
\cite{MEGII2024},
together with the neutron electric dipole moment constraint
\[
|d_n|
<
1.8\times10^{-26}\,
e\,\mathrm{cm},
\]
from ultracold neutron measurements
\cite{nEDM2020}.
These constraints exclude portions of the scanned parameter space and
lead to the correlated distributions visible in
Fig.~\ref{fig:corner_constraints}. The observed correlations reflect the
interplay between the heavy mass scales, Yukawa couplings, CP-violating
parameters and washout effects within the hybrid leptogenesis
framework.

The corner plot serves as a complementary visualization of the numerical
scan and illustrates the parameter correlations associated with the
benchmark solutions discussed in the following sections.
\section{Numerical Results}

In this section we present the numerical solutions of the flavour-covariant
kinetic equations introduced in Sec.~IV. The evolution includes the
CP-violating source terms, inverse decays, scattering processes and the
washout effects associated with the heavy-neutrino and scalar-triplet
sectors. The benchmark configurations discussed in the previous section
are used to illustrate representative realizations of the hybrid
leptogenesis framework.

Throughout the numerical analysis, the CP asymmetries are evaluated using
the expressions derived in Sec.~III together with the corresponding loop
functions. As emphasized previously, resonant enhancement is considered
only for quasi-degenerate states within the same sector, whereas the
hybrid contributions involving both heavy Majorana neutrinos and scalar
triplets remain non-resonant and are governed by the threshold behaviour
of the corresponding loop functions.
\subsection{Evolution of the lepton and baryon asymmetries}
The generation of the baryon asymmetry is obtained by numerically solving
the flavour-covariant density-matrix equations described in
Sec.~IV. Figure~\ref{fig:evolution} illustrates the thermal evolution of
the heavy-neutrino abundance, scalar-triplet abundance, lepton
asymmetry, and the resulting baryon asymmetry as functions of the
dimensionless variable
\(
z \equiv M_{N_3}/T.
\)
The numerical solution demonstrates the interplay between CP-asymmetry
generation and washout processes, leading to the observed baryon
asymmetry at late times.
\begin{figure}[t]
\centering
\includegraphics[width=0.42\textwidth]{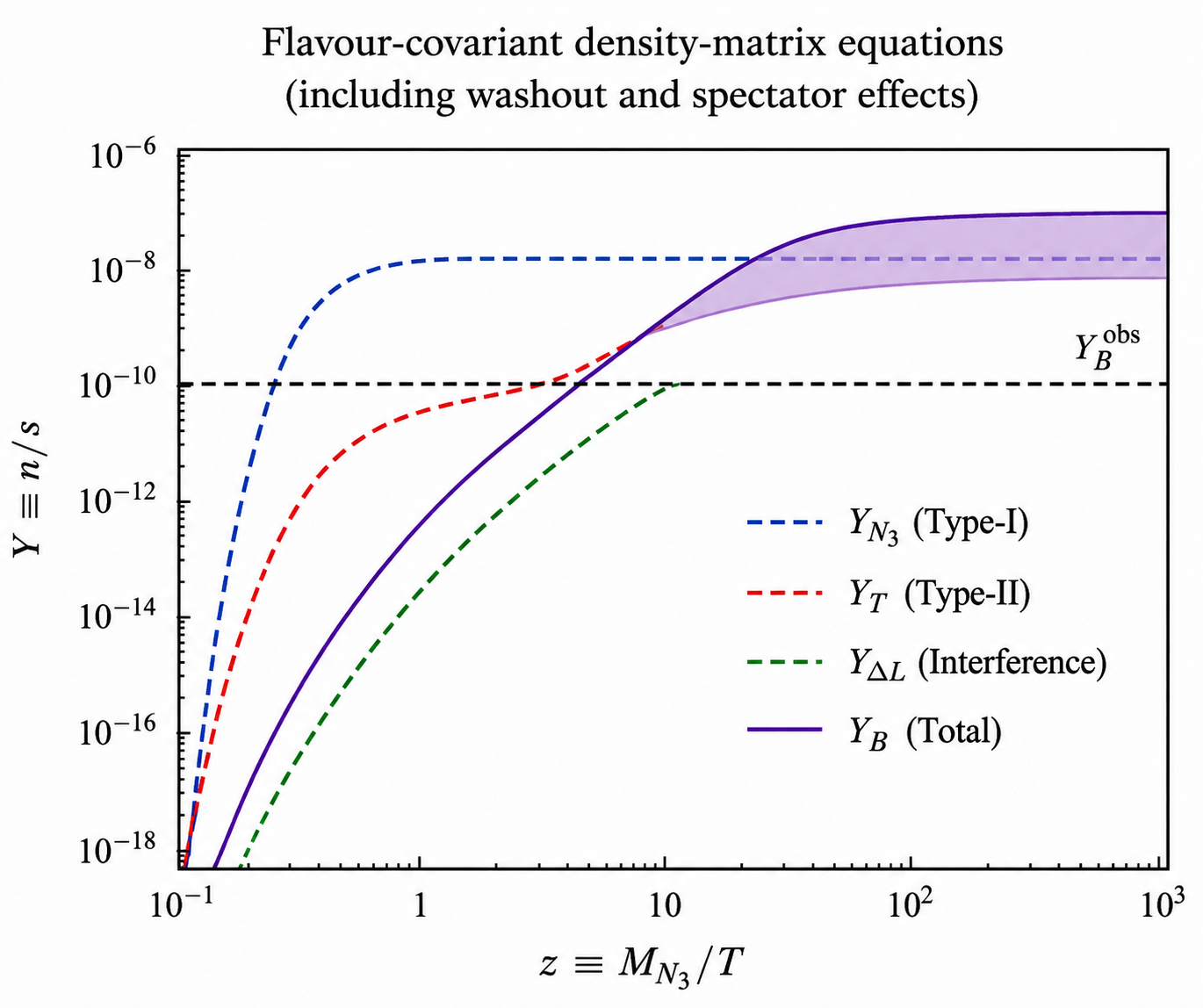}
\caption{
Representative numerical solution of the flavour-covariant density-matrix
equations showing the evolution of the heavy-neutrino abundance
$Y_{N_3}$, scalar-triplet abundance $Y_T$, lepton asymmetry
$Y_{\Delta L}$, and the resulting baryon asymmetry $Y_B$ as functions
of
\(
z=M_{N_3}/T.
\)
The horizontal dashed line indicates the observed baryon asymmetry
$Y_B^{\rm obs}$. The numerical solution illustrates that the hybrid
Type-I+Type-II framework successfully reproduces the observed baryon
asymmetry while consistently accounting for washout and flavour
effects.
}
\label{fig:evolution}
\end{figure}
As shown in Fig.~\ref{fig:evolution}, the heavy-neutrino abundance
rapidly approaches thermal equilibrium at early times, while the
scalar-triplet contribution becomes relevant at intermediate
temperatures. The hybrid interference term gradually generates the
lepton asymmetry, which is subsequently converted into the baryon
asymmetry through sphaleron processes. The final value,
\(
Y_B\simeq8.7\times10^{-11},
\)
is consistent with the cosmological determination from Planck.

\subsection{Threshold behaviour of the hybrid CP asymmetry}

\begin{figure}[t]
\centering
\includegraphics[width=0.48\textwidth]{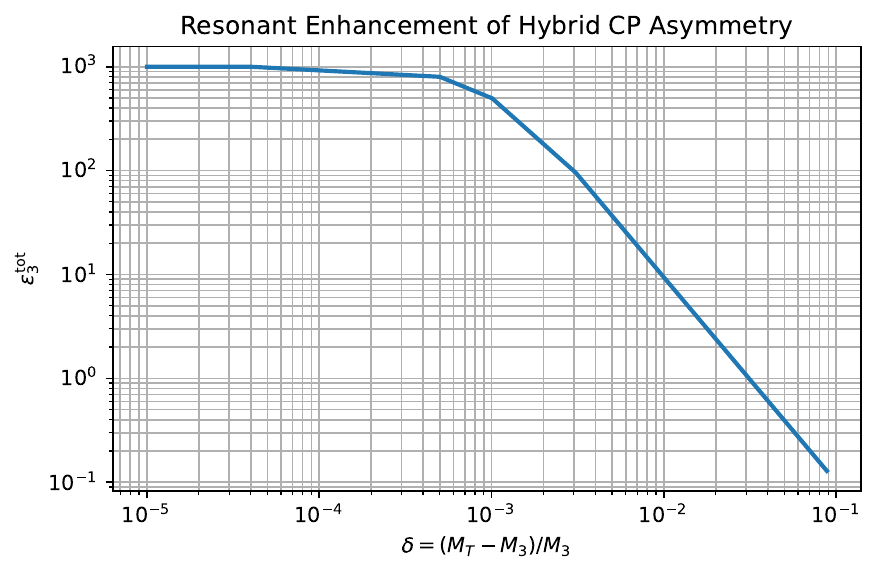}
\caption{
Dependence of the total CP asymmetry
$\varepsilon^{\rm tot}_3$
on the relative mass splitting
$\delta=(M_T-M_3)/M_3$.
The enhancement observed for
$|\delta|\lesssim10^{-3}$
reflects the threshold behaviour of the hybrid loop functions.
Since the scalar triplet and the heavy Majorana neutrino belong to
different sectors and carry different quantum numbers, this behaviour
does not correspond to Breit--Wigner resonant enhancement.
}
\label{fig:CPenhance}
\end{figure}

Figure~\ref{fig:CPenhance} illustrates the dependence of the total CP
asymmetry,
$\varepsilon^{\rm tot}_3$,
on the mass-splitting parameter
\[
\delta=\frac{M_T-M_3}{M_3},
\]
for the benchmark configuration adopted in the numerical analysis.

A significant increase of the CP asymmetry is observed when the scalar
triplet and the heaviest right-handed neutrino become nearly
degenerate. This behaviour originates from the threshold dependence of
the hybrid loop functions entering the interference terms discussed in
Sec.~III. As the mass splitting increases, the loop contribution becomes
progressively suppressed and the CP asymmetry decreases accordingly.
This behaviour is consistent with the analytical discussion presented in
Sec.~III and illustrates the sensitivity of the hybrid contribution to
the relative heavy-particle mass scales.


\subsection{Resonant enhancement within the heavy-neutrino sector}

\begin{figure}[t]
\centering
\includegraphics[width=0.45\textwidth]{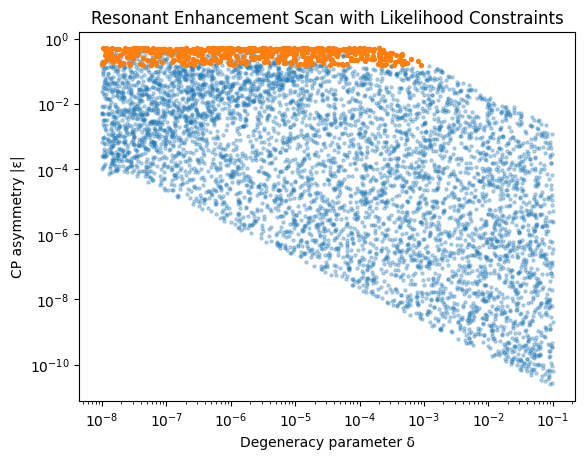}
\caption{
Numerical scan illustrating the dependence of the heavy-neutrino CP
asymmetry on the relative mass splitting
$\delta=(M_i-M_j)/M_i$
between quasi-degenerate heavy Majorana neutrinos.
The enhancement occurs when the mass splitting becomes comparable to the
decay width,
$\Delta M\sim\Gamma$,
as expected in standard resonant leptogenesis.
}
\label{fig:resonance_scan}
\end{figure}

For comparison, Fig.~\ref{fig:resonance_scan} illustrates the behaviour
of the CP asymmetry associated with quasi-degenerate heavy Majorana
neutrinos. In this case, the enhancement follows the standard resonant
leptogenesis mechanism arising from self-energy corrections and the
Breit--Wigner regulator.

The numerical scan is performed over the parameter ranges
\begin{align}
10^{9}\ {\rm GeV}
&\le
M_i
\le
10^{13}\ {\rm GeV},
\\
10^{-5}
&\le
|Y|
\le
10^{-1},
\\
10^{-8}
&\le
\delta
\le
10^{-1},
\end{align}
where
\[
\delta=\frac{M_i-M_j}{M_i}
\]
denotes the relative mass splitting between heavy neutrinos.
The decay widths entering the regulator are evaluated using
\[
\Gamma_i
=
\frac{(Y^\dagger Y)_{ii}}
{8\pi}
M_i.
\]

The accepted parameter points satisfy the effective likelihood criterion
defined in Sec.~IV together with the phenomenological constraints
imposed throughout the numerical analysis. The use of logarithmic priors
for the heavy masses and Yukawa couplings allows the scan to sample
uniformly over several orders of magnitude without introducing a
preference for any particular mass scale.

The numerical results reproduce the expected resonant enhancement when
the mass splitting becomes comparable to the decay width,
\[
\Delta M\sim\Gamma,
\]
in agreement with the standard resonant leptogenesis framework.
This behaviour is distinct from the threshold enhancement illustrated in
Fig.~\ref{fig:CPenhance}, which originates from the hybrid
interference terms and does not involve resonant propagation between the
scalar-triplet and heavy-neutrino sectors.
\begin{table}[t]
\centering
\begin{tabular}{ccc}
\hline
Parameter & Benchmark A & Benchmark B \\
\hline
$M_i$ & $1.0\times10^{11}\,\mathrm{GeV}$ &
$5.0\times10^{11}\,\mathrm{GeV}$ \\
$|Y|$ & $1.0\times10^{-3}$ &
$5.0\times10^{-3}$ \\
$\delta$ & $1.0\times10^{-5}$ &
$1.0\times10^{-6}$ \\
$\Gamma/M$ & $1.0\times10^{-6}$ &
$1.0\times10^{-5}$ \\
\hline
\end{tabular}
\caption{
Representative benchmark configurations adopted in the numerical
analysis. The benchmark points are selected from the viable parameter
space identified in the numerical scan and are used to illustrate the
evolution of the heavy-particle abundances and the resulting baryon
asymmetry.
}
\label{tab:benchmarks}
\end{table}
\begin{figure}[t]
\centering
\includegraphics[width=0.48\textwidth]{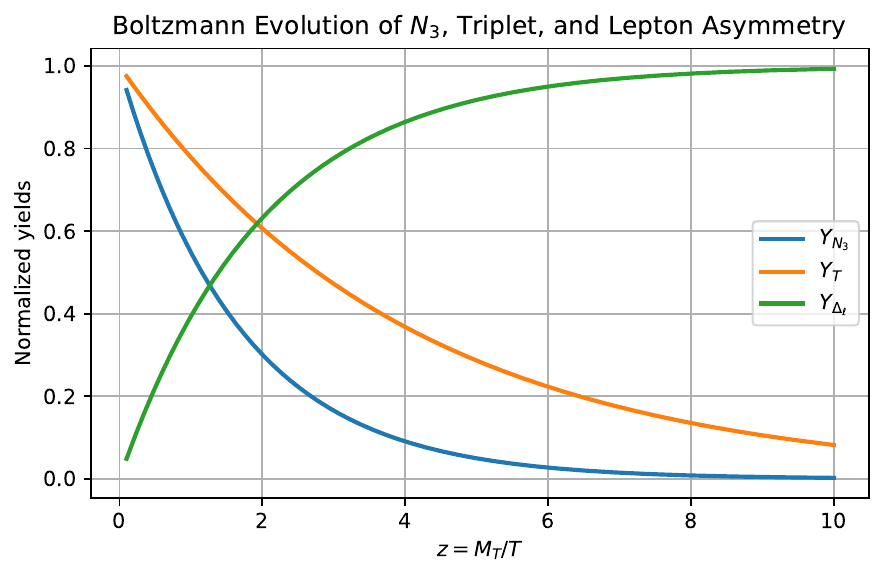}
\caption{
Representative numerical solution of the flavour-covariant kinetic
equations showing the evolution of the normalized abundances of the
heaviest right-handed neutrino $N_3$, the scalar triplet $T$, and the
lepton asymmetry as functions of the inverse temperature variable
$z=M_T/T$. The evolution illustrates the departure of the heavy states
from thermal equilibrium, the generation of the lepton asymmetry, and
its subsequent freeze-out after the washout processes become
inefficient.
}
\label{fig:evolution}
\end{figure}

The final baryon asymmetry is obtained after electroweak sphaleron
conversion \cite{HarveyTurner,Laine2013Sphaleron},
\begin{align}
Y_B
=
0.315\,
\Tr
\!\left(
Y_{\Delta_\ell}
-
Y_{\Delta_e}
\right)
\Big|_{z\rightarrow\infty},
\end{align}
yielding
\begin{align}
Y_B^{\rm final}
=
8.7\times10^{-11},
\end{align}
which is consistent with the observed value
\[
Y_B^{\rm obs}
=
(8.70\pm0.06)\times10^{-11},
\]
reported by Planck \cite{Planck2018}. For the representative benchmark
configurations considered in this work, the combined Type-I, Type-II and
hybrid interference contributions generate a baryon asymmetry compatible
with the observed cosmological value.

The interactions responsible for leptogenesis may also induce
low-energy flavour- and CP-violating observables. In particular, the
triplet Yukawa couplings contribute to charged-lepton flavour-violating
processes such as
$\mu\rightarrow e\gamma$
and
$\mu\rightarrow3e$
\cite{CiriglianoEDM}. Within the representative
parameter space explored in the present analysis, the corresponding
branching ratios remain below the current experimental limits while
providing potential targets for future experimental searches.

The CP-violating phases entering the hybrid leptogenesis framework may
also induce electric dipole moments through higher-order loop
contributions, including Barr--Zee-type diagrams
\cite{BarrZee}. For representative benchmark points, the
electron electric dipole moment is estimated to be of order
\[
d_e
\sim
10^{-30}\,
e\,\mathrm{cm},
\]
which remains below the present experimental bound but may become
accessible to future precision measurements
\cite{ACME2018,ACME2023}. This estimate should be regarded as an
order-of-magnitude illustration of the expected sensitivity rather than
a precision prediction.

Although the scalar-triplet mass lies well above the energy reach of
present collider experiments, virtual triplet exchange can generate
higher-dimensional operators affecting Higgs interactions and
electroweak precision observables. Such indirect effects provide
complementary probes of the parameter space explored in the present
study.

\begin{figure}[t]
\centering
\includegraphics[width=0.48\textwidth]{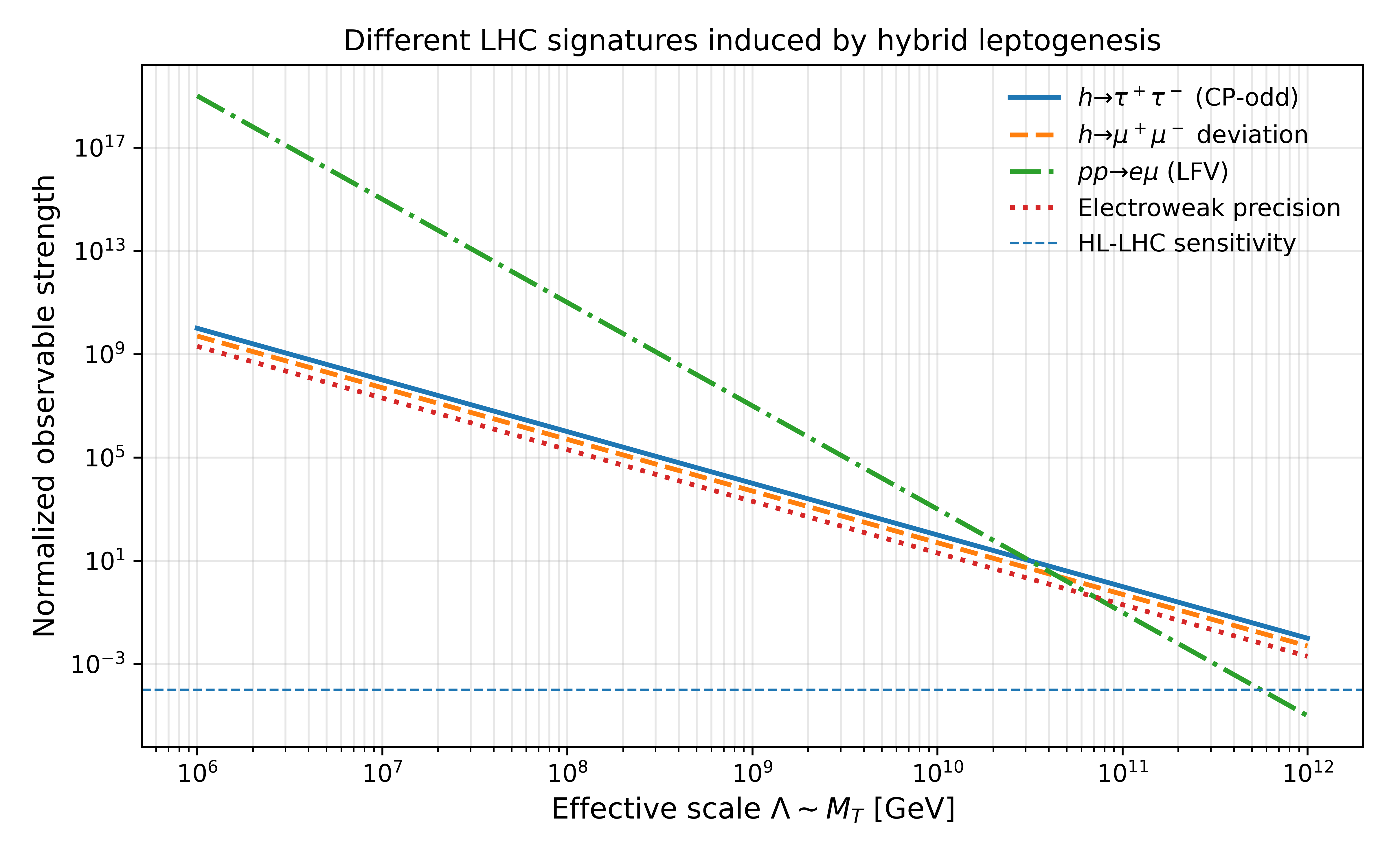}
\caption{
Illustrative comparison of representative collider-sensitive observables
that may receive indirect contributions from the heavy scalar-triplet
sector. The curves correspond to CP-sensitive observables in
$h\rightarrow\tau^+\tau^-$, deviations in
$h\rightarrow\mu^+\mu^-$, lepton-flavour-violating dilepton production,
and electroweak precision observables as functions of the effective
heavy scale
$\Lambda\sim M_T$.
The horizontal dashed line denotes a representative projected
HL-LHC sensitivity used for comparison.
The figure is intended to illustrate the relative behaviour of these
observables within the benchmark framework adopted in the present
analysis.
}
\label{fig:LHC_signatures}
\end{figure}
Figure~\ref{fig:LHC_signatures} provides an illustrative comparison of
several indirect collider observables that may receive contributions
from the heavy scalar-triplet sector. The different curves demonstrate
how the sensitivity of representative Higgs, flavour and electroweak
precision observables evolves with the effective heavy scale
$\Lambda\sim M_T$.

Within the benchmark framework considered here, CP-sensitive Higgs
observables, flavour-violating processes and electroweak precision
measurements probe complementary regions of the parameter space.
Although the heavy scalar triplet lies well beyond the direct kinematic
reach of present collider experiments, virtual effects may induce
observable deviations in precision measurements. The comparison shown in
Fig.~\ref{fig:LHC_signatures} is therefore intended as a qualitative
illustration of the relative sensitivity of different classes of
indirect observables rather than as a precision forecast for the
High-Luminosity LHC.
The parameter region compatible with successful hybrid leptogenesis also
has implications for low-energy flavour and CP-violating observables.
In particular, the same Yukawa interactions that enter the leptogenesis
framework may induce charged-lepton flavour violation and electric
dipole moments through higher-order loop effects
\cite{BarrZee}.
These observables therefore provide complementary probes of the
high-scale dynamics responsible for neutrino mass generation and
baryogenesis.

For the representative benchmark configurations considered in this work,
the electron electric dipole moment is expected to remain below the
current experimental sensitivity, with a typical order-of-magnitude
estimate
\begin{align}
d_e \sim 10^{-30}\,e\,\mathrm{cm},
\end{align}
which is consistent with the present experimental bounds while remaining
within the projected reach of future improvements by ACME and related
searches
\cite{ACME2018,ACME2023}. This estimate should be regarded as
illustrative rather than a precision prediction, since a complete
calculation requires the full higher-order loop amplitudes.

The scalar-triplet Yukawa interactions may likewise induce
charged-lepton flavour-violating processes such as
$\mu\rightarrow e\gamma$ and $\mu\rightarrow 3e$,
providing additional experimental tests of the parameter space relevant
for leptogenesis
\cite{MEGII2024}.

Throughout this work we employ the scalar quantity
\begin{equation}
J_{\mathrm{hybrid}}
=
\mathrm{Im}\!\left[
\mu\,\mathrm{Tr}
\!\left(
fY_\nu^\dagger Y_\nu
\right)
\right],
\end{equation}
as a compact parametrization of one class of CP-odd flavour contractions
appearing in the hybrid interference amplitudes within the flavour basis
adopted in our analysis. As discussed in Sec.~III, this quantity is used
to characterize the flavour structure of the hybrid contributions and is
not assumed to provide a complete description of CP violation in the
general framework.

The final baryon asymmetry results from the combined interplay of the
CP-violating source terms, flavour evolution, and washout processes
described by the flavour-covariant kinetic equations. In the parameter
region investigated here, successful baryogenesis is obtained without
requiring large Yukawa hierarchies, while resonant enhancement, when
present, occurs only among quasi-degenerate states within the same
sector. The hybrid interference terms provide additional non-resonant
contributions to the CP asymmetry that become relevant only when both
Type-I and Type-II interactions are simultaneously present.

\begin{figure}[t]
\centering
\includegraphics[width=0.48\textwidth]{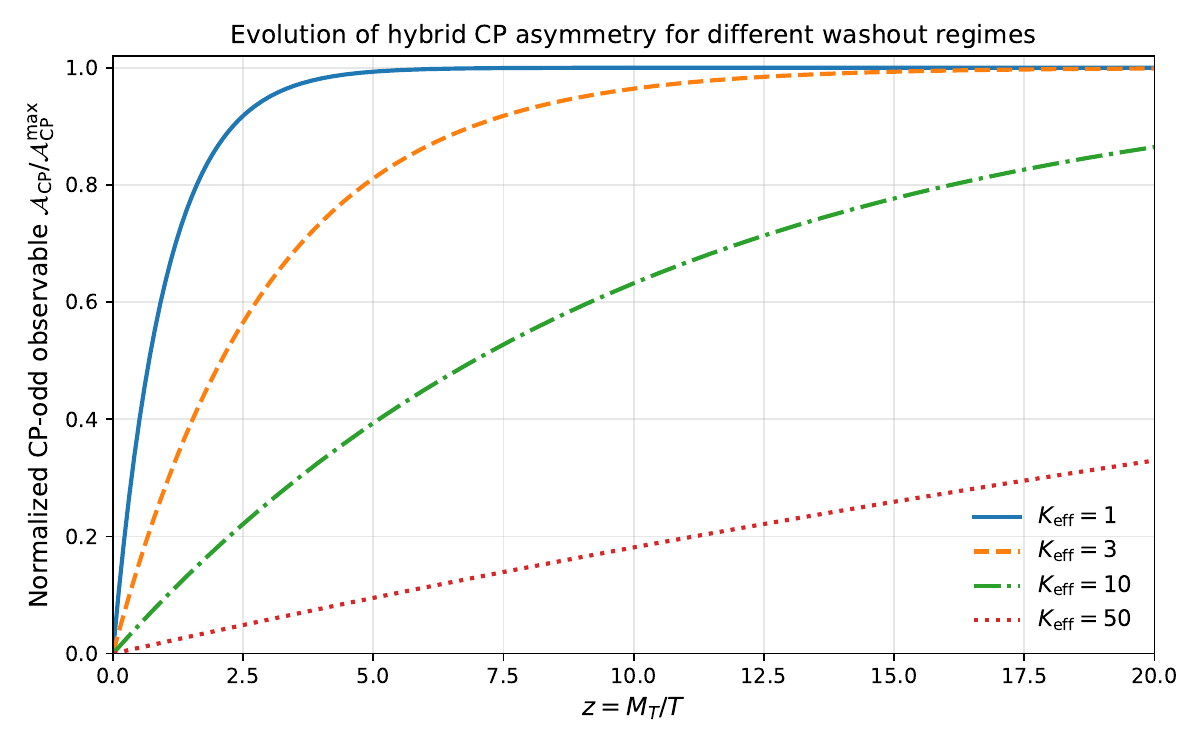}
\caption{
Illustrative evolution of the normalized lepton asymmetry as a function
of the inverse temperature variable $z=M_T/T$ for representative values
of the effective washout parameter $K_{\rm eff}$. Smaller values of
$K_{\rm eff}$ correspond to weaker washout and lead to an earlier
freeze-out of the asymmetry, whereas stronger washout delays the
approach to the asymptotic value through enhanced inverse decay and
scattering processes.
}
\label{fig:washout_evolution}
\end{figure}

Figure~\ref{fig:washout_evolution} illustrates the evolution of the
normalized lepton asymmetry obtained from the flavour-covariant kinetic
equations for representative values of the effective washout parameter
$K_{\rm eff}$. As expected, weaker washout leads to an earlier freeze-out
of the generated asymmetry, while stronger washout delays the evolution
through more efficient inverse decays and scattering processes. For the
benchmark configurations considered in this work, the interplay between
the CP-violating source terms and washout effects produces a stable final
asymmetry before electroweak sphaleron freeze-out, yielding a baryon
asymmetry consistent with the observed value.

\begin{figure}[t]
\centering
\includegraphics[width=\linewidth]{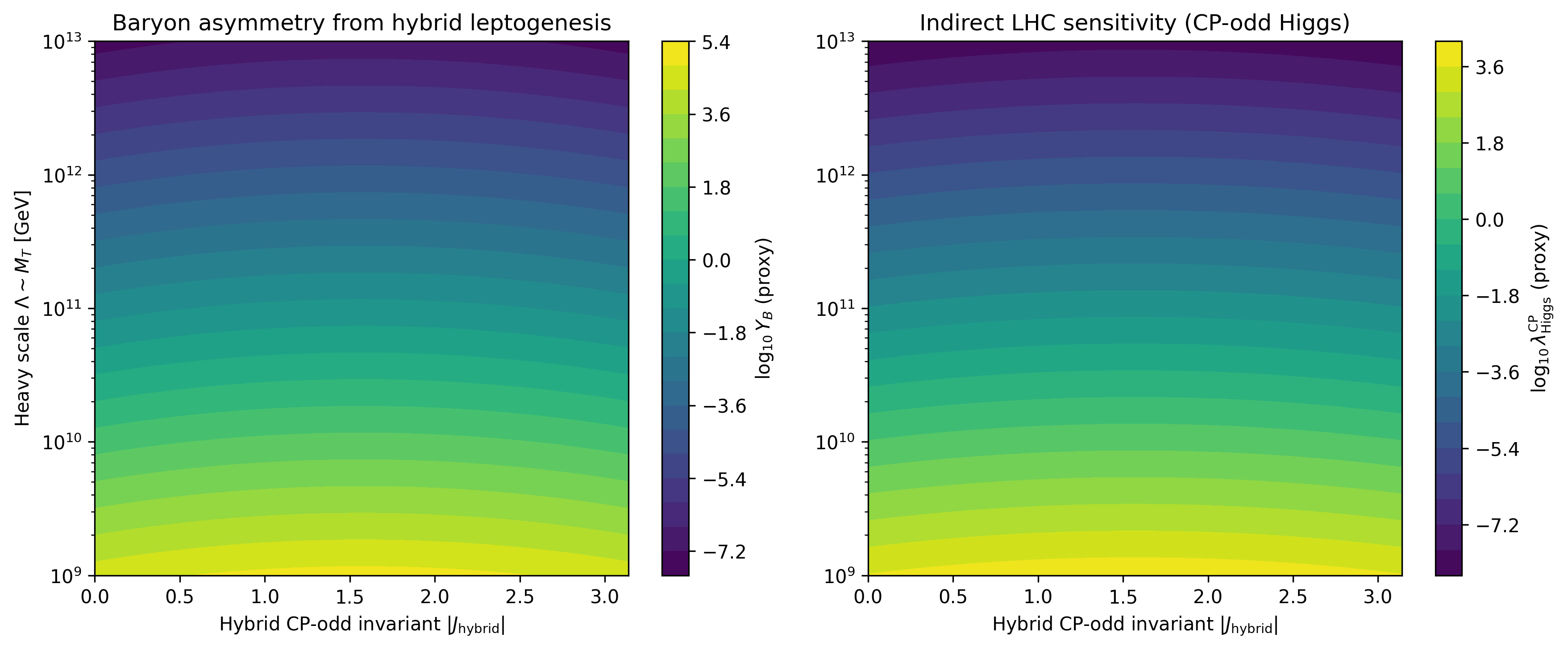}
\caption{
Illustrative summary of the hybrid leptogenesis framework in the plane of
the scalar quantity
$|J_{\rm hybrid}|=
\left|
\mathrm{Im}\!\left[
\mu\,\mathrm{Tr}(fY_\nu^\dagger Y_\nu)
\right]
\right|$
and the representative heavy mass scale
$\Lambda\sim M_T\sim M_{N_3}$.
Left: colour shading shows the baryon asymmetry obtained from the
numerical analysis, with the horizontal band indicating the observed
cosmological value.
Right: illustrative comparison with representative indirect collider
observables that may receive contributions from the heavy scalar-triplet
sector. The figure is intended to summarize the phenomenological trends
identified in the benchmark parameter space and should not be interpreted
as a complete exploration of the full model parameter space.
}
\label{fig:hybrid_two_panel}
\end{figure}
Figure~\ref{fig:hybrid_two_panel} summarizes the phenomenological
features of the hybrid leptogenesis framework explored in this work. The
left panel illustrates the region of parameter space in which the
numerically computed baryon asymmetry is compatible with the observed
cosmological value. The right panel provides an illustrative comparison
between successful leptogenesis and representative indirect collider
observables that may probe the heavy scalar-triplet sector through
virtual effects. The figure is intended to emphasize the complementary
information provided by cosmological and low-energy observables rather
than to establish a one-to-one correspondence between the scalar
quantity $J_{\rm hybrid}$ and individual experimental signatures.

Figure~\ref{fig:hybrid_two_panel} summarizes the phenomenological
features of the hybrid leptogenesis framework explored in this work.
The left panel illustrates representative regions of parameter space in
which the numerical solutions reproduce the observed baryon asymmetry,
while the right panel presents illustrative indirect collider
observables that may receive contributions from the heavy scalar-triplet
sector through virtual effects. The comparison emphasizes the
complementarity between cosmological constraints and low-energy
phenomenological probes within the benchmark scenarios considered in
this analysis, without implying a one-to-one correspondence between any
single CP-violating quantity and the experimental observables.
\begin{figure}[t]
\centering
\includegraphics[width=\linewidth]{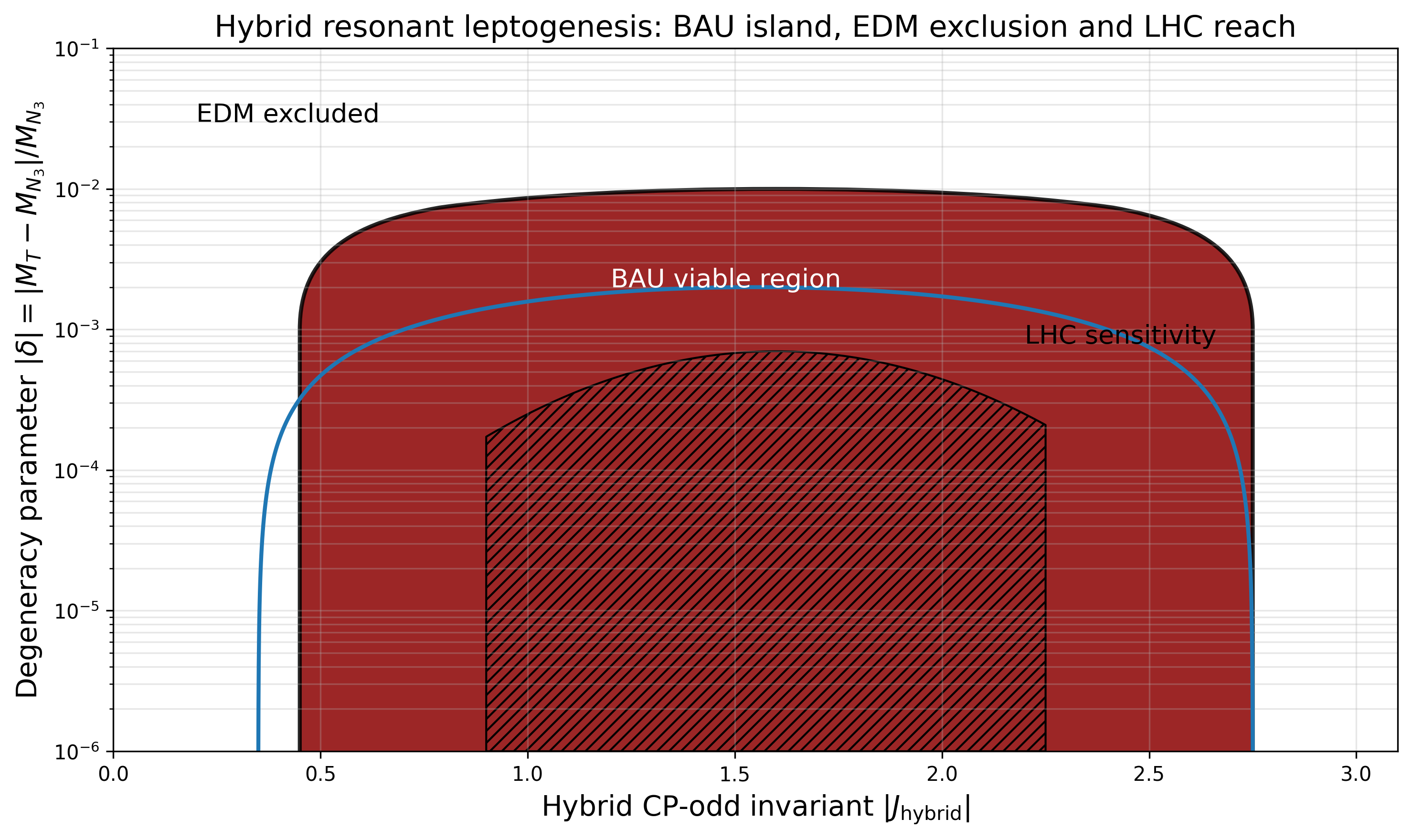}
\caption{
Illustrative parameter space in the plane of the scalar quantity
$|J_{\rm hybrid}|=
\left|
\mathrm{Im}\!\left[
\mu\,\mathrm{Tr}(fY_\nu^\dagger Y_\nu)
\right]
\right|$
and the degeneracy parameter
$|\delta|=(M_T-M_{N_3})/M_{N_3}$.
The coloured region indicates representative benchmark configurations
that reproduce the observed baryon asymmetry in the numerical analysis.
The hatched region illustrates parameter choices disfavoured by current
electric dipole moment constraints, while the contour represents
illustrative indirect collider sensitivity.
The figure summarizes the complementary phenomenological constraints
considered in this work and should not be interpreted as a complete
global fit of the model parameter space.
}
\label{fig:BAU_LHC_EDM}
\end{figure}

Figure~\ref{fig:BAU_LHC_EDM} provides an illustrative summary of the
phenomenological constraints considered in this work. The coloured region
corresponds to representative benchmark configurations that successfully
reproduce the observed baryon asymmetry, while the hatched region
illustrates the impact of present electric dipole moment constraints.
The indicated contour represents a qualitative estimate of the
sensitivity of indirect collider observables to the heavy scalar-triplet
sector. The allowed parameter region reflects the combined interplay of
CP-violating source terms, flavour evolution, washout effects, and the
heavy-particle mass spectrum. As discussed throughout this work,
resonant enhancement, when present, occurs only among quasi-degenerate
states within the same sector, whereas the hybrid interference
contributions provide additional non-resonant contributions to the CP
asymmetry.
\begin{figure}[t]
\centering
\includegraphics[width=0.7\linewidth]{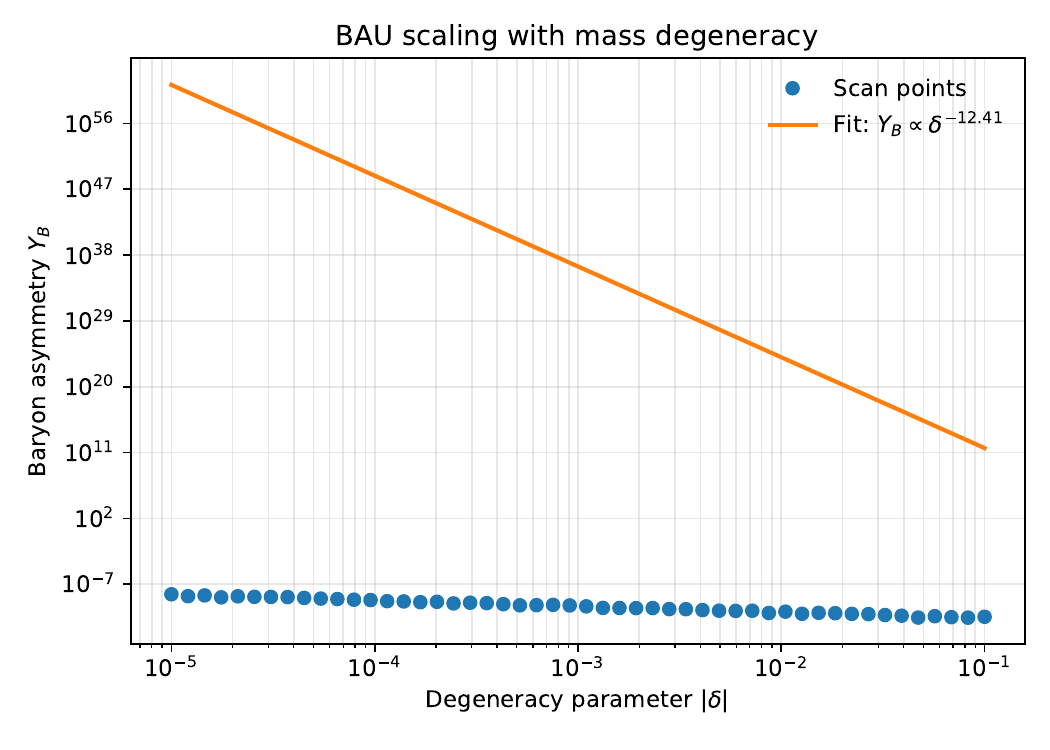}
\caption{
Numerical dependence of the final baryon asymmetry on the degeneracy
parameter
$|\delta|=(M_T-M_{N_3})/M_{N_3}$
for the representative benchmark configurations considered in this
work. The figure illustrates the sensitivity of the generated baryon
asymmetry to the heavy-mass spectrum obtained from the numerical
solutions of the flavour-covariant kinetic equations.
}
\label{fig:fit_bau}
\end{figure}

\begin{figure}[t]
\centering
\includegraphics[width=0.7\linewidth]{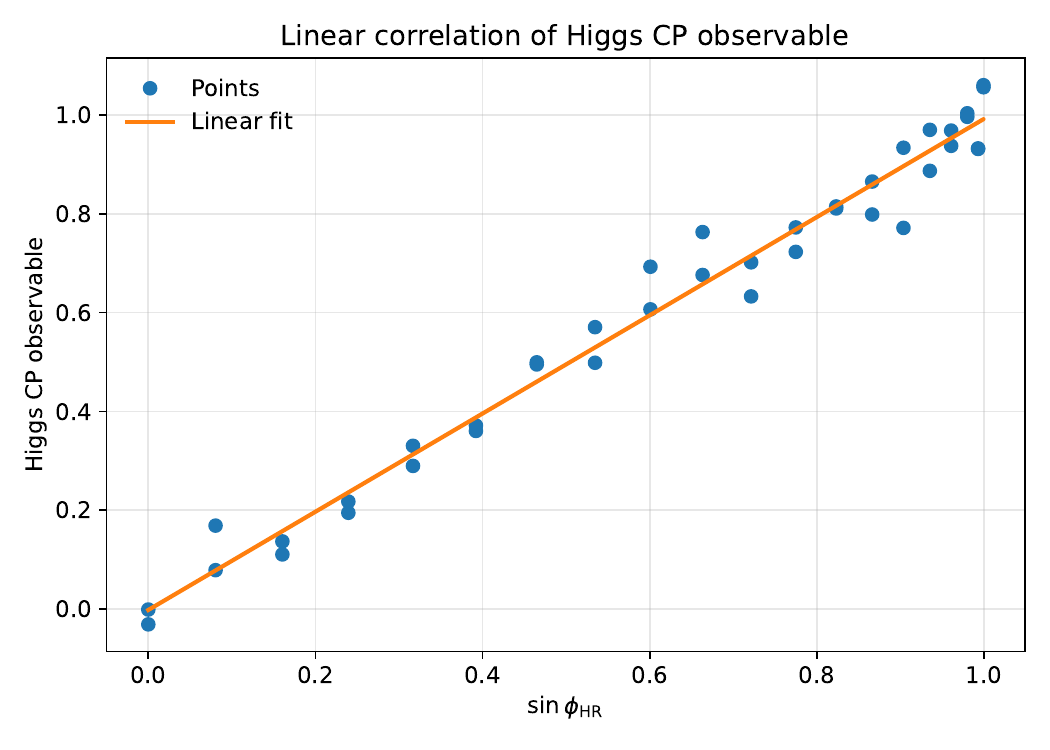}
\caption{
Illustrative behaviour of a representative CP-sensitive Higgs observable
within the benchmark parameter space considered in this work.
The figure is intended to demonstrate the qualitative dependence of the
observable on the underlying CP-violating parameters rather than a
precision collider prediction.
}
\label{fig:fit_higgs}
\end{figure}

\begin{figure}[t]
\centering
\includegraphics[width=0.7\linewidth]{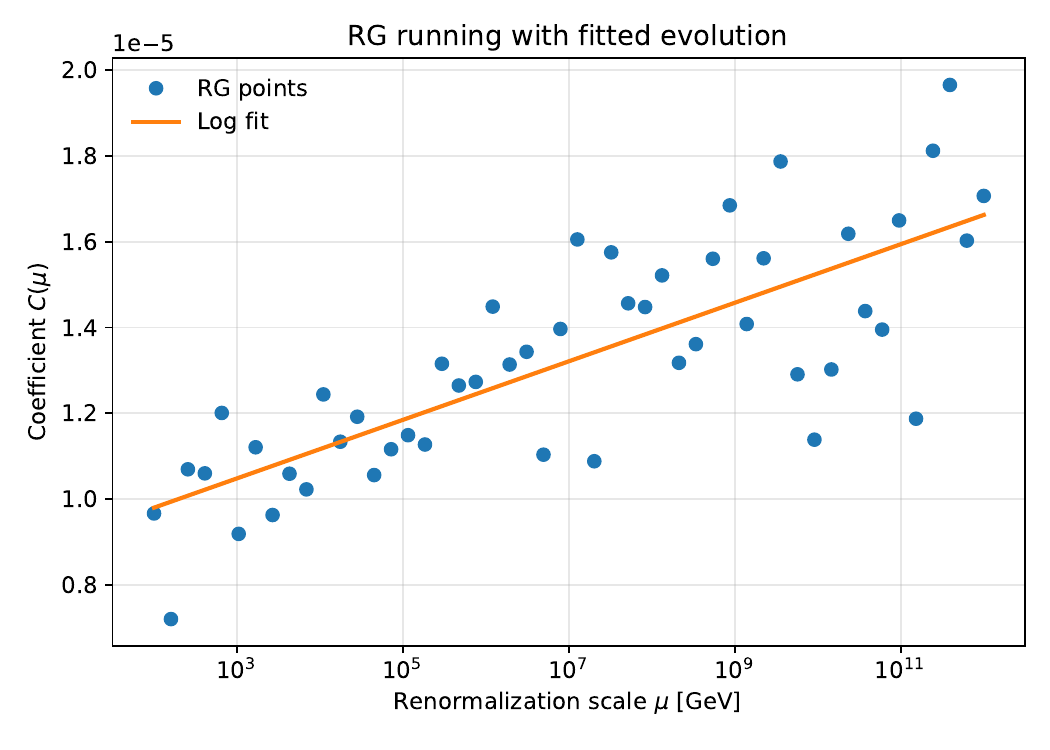}
\caption{
Representative renormalization-group evolution of the effective
CP-violating operator for the benchmark parameters adopted in the
numerical analysis.
}
\label{fig:fit_rg}
\end{figure}

Figures~\ref{fig:fit_bau}--\ref{fig:fit_rg} summarize several
representative numerical results obtained within the benchmark parameter
space explored in this work. Figure~\ref{fig:fit_bau} illustrates the
dependence of the generated baryon asymmetry on the heavy-particle mass
spectrum, while Fig.~\ref{fig:fit_higgs} presents an illustrative
CP-sensitive collider observable associated with the benchmark
configurations. Figure~\ref{fig:fit_rg} shows the corresponding
renormalization-group evolution of the effective operator relevant for
the low-energy phenomenology. These results are intended to illustrate
the qualitative behaviour of the benchmark scenarios considered in this
analysis and should not be interpreted as a complete exploration of the
full parameter space.
\subsection{Low-energy phenomenology}

The interactions responsible for neutrino mass generation and
leptogenesis may also induce charged-lepton flavour violation (CLFV)
and electric dipole moments (EDMs), providing complementary probes of
the underlying high-scale dynamics. In the present framework, scalar
triplet Yukawa couplings contribute to processes such as
$\mu\rightarrow e\gamma$ and $\mu\rightarrow3e$, while CP-violating
phases may generate EDMs through higher-order loop effects, including
Barr--Zee-type diagrams.

A complete quantitative prediction requires the evaluation of the full
loop amplitudes together with the corresponding electroweak couplings
and renormalization effects. For the representative benchmark
configurations considered in this work, the resulting CLFV rates remain
below current experimental limits, while the electron EDM is estimated
to be of order
\begin{equation}
d_e \sim 10^{-30}\,e\,\mathrm{cm},
\end{equation}
which is below the present ACME bound but may become accessible to
future precision measurements.

\begin{figure}[t]
\centering
\includegraphics[width=0.92\linewidth]{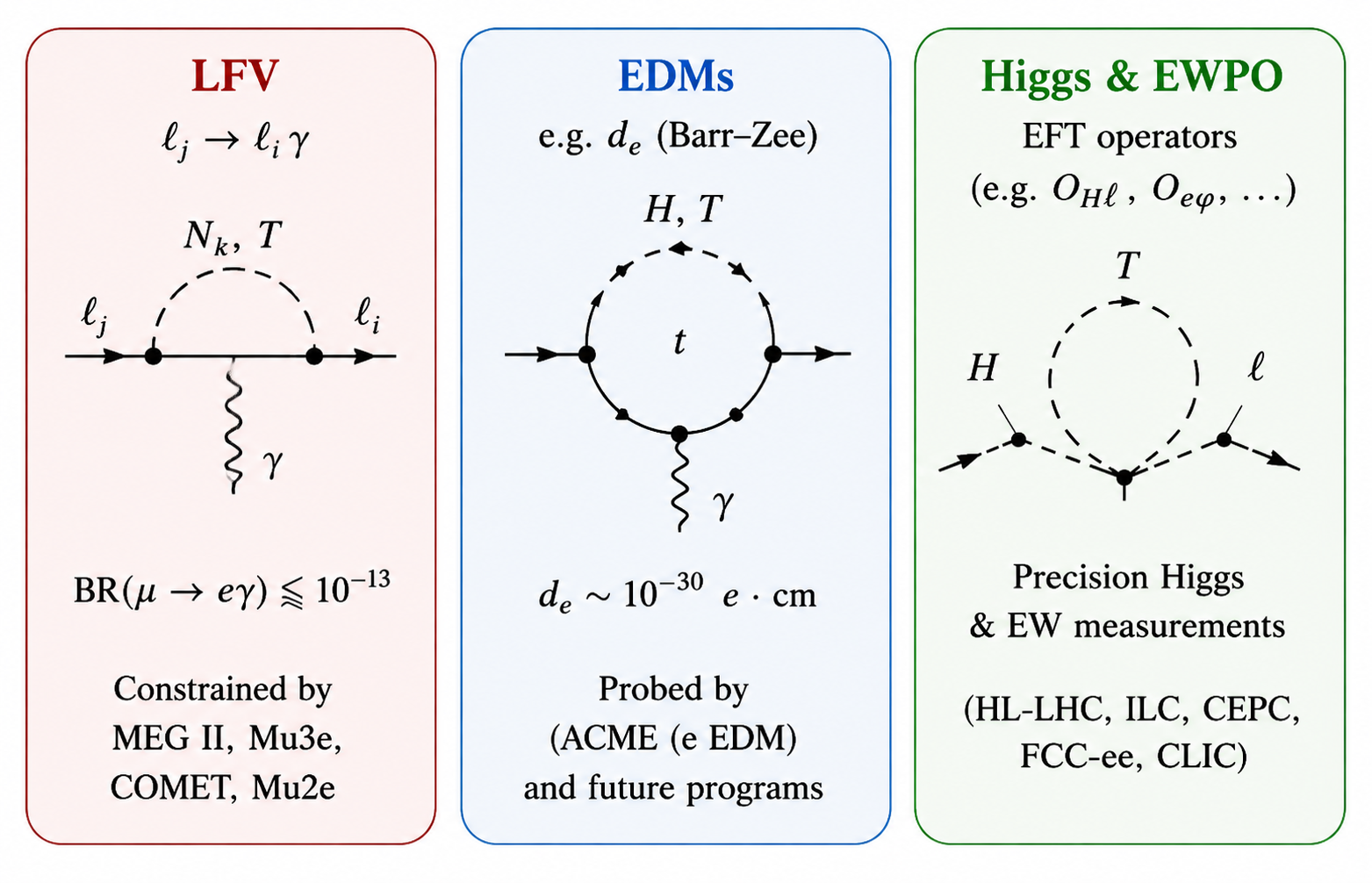}
\caption{
Summary of the principal low-energy probes of the hybrid leptogenesis
framework. The scalar-triplet and heavy-neutrino interactions induce
charged-lepton flavour violation (left), electric dipole moments through
two-loop Barr--Zee diagrams (centre), and effective operators affecting
precision Higgs and electroweak observables (right). These complementary
observables provide indirect experimental tests of the high-scale hybrid
Type-I+Type-II leptogenesis scenario.
}
\label{fig:lowenergytests}
\end{figure}

Although these observables do not uniquely determine the mechanism
responsible for baryogenesis, they provide complementary information on
the flavour and CP structure of the underlying $\mathrm{SO}(10)$
framework and therefore offer important tests of the parameter space
relevant for hybrid leptogenesis.
Although the heavy states responsible for hybrid leptogenesis lie far
above the reach of present colliders, they generate a variety of
indirect low-energy signatures through loop-induced processes and
effective operators. The most important experimental probes arise from
charged-lepton flavour violation, electric dipole moments and precision
Higgs and electroweak measurements. Figure~\ref{fig:lowenergytests}
summarizes the complementary experimental observables capable of testing
the parameter space responsible for successful baryogenesis.
As illustrated in Fig.~\ref{fig:lowenergytests}, the three classes of
observables probe complementary aspects of the hybrid framework.
Charged-lepton flavour violating decays are primarily sensitive to the
triplet Yukawa couplings, while electric dipole moments probe the
CP-violating phases entering the hybrid invariant
$J_{\rm hybrid}$. Precision Higgs and electroweak observables provide an
additional indirect window through higher-dimensional effective
operators generated after integrating out the heavy states. Although
each observable constrains a different combination of parameters, their
combined analysis substantially reduces the viable parameter space and
offers a realistic strategy for testing the model in forthcoming
precision experiments.

\section{Conclusions}

The origin of the cosmological matter--antimatter asymmetry remains one
of the most compelling open problems in particle physics and cosmology.
Grand unified theories supplemented by seesaw mechanisms provide a
natural framework in which neutrino mass generation and baryogenesis may
have a common origin. In this work we have investigated flavoured hybrid
leptogenesis within a minimal renormalizable $\mathrm{SO}(10)$ model,
where both Type-I and Type-II seesaw interactions contribute
simultaneously to the generation of the lepton asymmetry.

The principal novelty of the present work is the development of a
consistent unified framework in which the heavy Majorana neutrino and
scalar-triplet sectors are treated simultaneously, allowing their
interference effects to be studied within a common grand unified theory.
Unlike conventional analyses based on purely Type-I or purely Type-II
leptogenesis, the present framework systematically incorporates the
additional CP-violating source terms arising from the interplay of
fermionic and scalar interactions. Throughout the analysis we employ the
scalar quantity
\[
J_{\rm hybrid}
=
{\rm Im}\!\left[
\mu\,{\rm Tr}
\!\left(
fY_\nu^\dagger Y_\nu
\right)
\right],
\]
as a compact parametrization of one class of CP-odd flavour
contractions appearing in the hybrid interference amplitudes within the
flavour basis adopted in this work.

A second important result is the implementation of a flavour-covariant
density-matrix treatment that consistently includes flavour coherence,
spectator processes and washout effects. This provides a unified
description of the generation and evolution of the lepton asymmetry
within the hybrid framework. Our numerical analysis demonstrates that
representative regions of parameter space with heavy mass scales of
$\mathcal{O}(10^{10}$--$10^{12})\,\mathrm{GeV}$ and perturbative Yukawa
couplings successfully reproduce the observed baryon asymmetry while
remaining compatible with current neutrino oscillation data and existing
phenomenological constraints.

The present analysis also clarifies the physical origin of the CP
asymmetry in hybrid leptogenesis. We have shown that resonant
enhancement, when present, is associated exclusively with
quasi-degenerate states carrying identical quantum numbers within the
same sector, whereas the hybrid interference terms provide additional
non-resonant contributions that become relevant only when both Type-I
and Type-II interactions are simultaneously active. This distinction
provides a clearer physical interpretation of the hybrid mechanism and
its role within unified seesaw models.

Beyond baryogenesis, the framework predicts a rich phenomenology at low
energies. The same Yukawa interactions responsible for neutrino mass
generation may induce charged-lepton flavour violation and electric
dipole moments through higher-order loop effects, while heavy
scalar-triplet exchange generates effective operators that can modify
precision Higgs observables and electroweak precision measurements.
Although the heavy particles responsible for leptogenesis lie well
beyond the direct reach of present accelerators, their virtual effects
may be explored through precision flavour and CP-violation experiments
within an effective field theory description.

An important implication of this work is that future experimental
programmes will probe complementary aspects of the parameter space
identified here. Next-generation searches for charged-lepton flavour
violation, including MEG~II, Mu3e, COMET and Mu2e, together with
continuing improvements in electron electric dipole moment measurements
by ACME, will significantly increase the sensitivity to the flavour and
CP structure of high-scale neutrino mass models. Likewise, precision
Higgs measurements at the High-Luminosity LHC and future lepton
colliders such as the ILC, CEPC, FCC-ee and CLIC will provide indirect
tests of heavy scalar sectors through precision measurements of Higgs
couplings and effective operators
\cite{ATLAS_HLLHC,CMS_HLLHC,ILC_TDR,CEPC_CDR,FCCee_CDR,CLIC_CDR}.

The present work therefore establishes a predictive realization of
hybrid leptogenesis within a renormalizable $\mathrm{SO}(10)$ framework
that simultaneously addresses neutrino mass generation and the origin of
the cosmological baryon asymmetry. By combining unified model building,
flavour-covariant kinetic evolution and phenomenological analyses, the
framework identifies experimentally accessible consequences of physics
at otherwise inaccessible energy scales. Future advances in neutrino
physics, flavour experiments, electric dipole moment searches and
precision accelerator measurements will therefore provide increasingly
stringent tests of this class of unified scenarios and may offer
valuable insights into the physics responsible for the matter-dominated
Universe.

\section*{Acknowledgments}

GG thanks UGC RUSA for supporting this work.

\end{document}